# The same strain of *Piscine orthoreovirus* (PRV-1) is involved with the development of different, but related, diseases in Atlantic and Pacific Salmon in British Columbia

Emiliano Di Cicco[1,2]*, Hugh W. Ferguson[3], Karia H. Kaukinen[1], Angela D. Schulze[1], Shaorong Li[1],

Amy Tabata[1], Oliver P. Günther[4], Gideon Mordecai[5], Curtis A. Suttle[5,6], and Kristina M. Miller[1]*

[1] Pacific Biological Station, Fisheries and Oceans Canada, Nanaimo, BC, Canada

[2] Pacific Salmon Foundation, Vancouver, BC, Canada

[3] School of Veterinary Medicine, St. George's University, True Blue, Grenada, W. Indies

[4] Günther Analytics, Vancouver, BC, Canada

[5] Department of Earth, Ocean and Atmospheric Sciences, University of British Columbia, Vancouver, BC

[6] Department of Microbiology and Immunology, Department of Botany, Institute of Oceans and Fisheries, University of British Columbia, Vancouver, BC, Canada

***Corresponding authors**

**Email: Emiliano.DiCicco@dfo-mpo.gc.ca (ED)**

**Kristi.Saunders@dfo-mpo.gc.ca (KMM)**

# Abstract

*Piscine orthoreovirus* Strain PRV-1 is the causative agent of heart and skeletal muscle inflammation (HSMI) in Atlantic salmon (*Salmo salar*). Given its high prevalence in net pen salmon, debate has arisen on whether PRV poses a risk to migratory salmon, especially in British Columbia (BC) where commercially important wild Pacific salmon are in decline. Various strains of PRV have been associated with diseases in Pacific salmon, including erythrocytic inclusion body syndrome (EIBS), HSMI-like disease, and jaundice/anemia in Japan, Norway, Chile and Canada. We examine the developmental pathway of HSMI and jaundice/anemia associated with PRV-1 in farmed Atlantic and Chinook (*Oncorhynchus tshawytscha*) salmon in BC, respectively. *In situ* hybridization localized PRV-1 within developing lesions in both diseases. The two diseases showed dissimilar pathological pathways, with inflammatory lesions in heart and skeletal muscle in Atlantic salmon, and degenerative-necrotic lesions in kidney and liver in Chinook salmon, plausibly explained by differences in PRV load tolerance in red blood cells. Viral genome sequencing revealed no consistent differences in PRV-1 variants intimately involved in the development of both diseases, suggesting that migratory Chinook salmon may be at more than a minimal risk of disease from exposure to the high levels of PRV occurring on salmon farms.

# Introduction

Viruses in the species *Piscine orthoreovirus* (PRV) were first discovered in association with farmed Atlantic salmon (*Salmo salar*) diagnosed with Heart and Skeletal Muscle Inflammation (HSMI) (Palacios et al. 2010), although HSMI had been reported within the Norwegian farming industry since 1999 (Kongtorp et al. 2004a). Even though the disease was infectious (Kongtorp and Taksdal 2009), no virus was cultured from affected fish (Kongtorp et al. 2004a,b; Watanabe et al. 2006; Kongtorp and Taksdal 2009). Moreover, disease challenge studies using viral filtrates from tissue homogenates of Atlantic salmon with HSMI (Kongtorp and Taksdal 2009) replicated the patterns of inflammation in heart and skeletal tissue that are diagnostic of this disease (Biering and Garseth 2012). However, only recently was a cause-and-effect relationship established through a challenge based on an isolate (NOR2012-V3621) of *Piscine orthoreovirus* purified from blood (Wessel et al. 2017), the primary target for the virus (Finstad et al. 2014, Wessel et al. 2015).

Since its discovery in 1999, the prevalence of HSMI has expanded from <30 farms in central Norway to hundreds of farms throughout the country (Bornø and Lie Linaker 2015). While HSMI became a reportable disease in 2004 (https://lovdata.no/dokument/SF/forskrift/2008-06-17-819), it is no longer reportable due both to the ubiquitous distribution of *Piscine orthoreovirus* (strains of which are collectively referred to as PRV) and the large-scale distribution of HSMI outbreaks (Bornø and Lie Linaker 2015). Moreover, while HSMI was initially reported only in ocean net pens, generally 5 to 9 months post-stocking (Kongtorp et al. 2004a), there are recent reports of HSMI outbreaks occurring in Norwegian freshwater hatcheries (Olsen et al. 2015; Hjeltnes et al. 2016, 2017). HSMI has also been reported in farmed Atlantic salmon in Scotland and the United Kingdom (Ferguson et al. 2005), Chile (Bustos et al. 2011; Godoy et al. 2016) and Canada (Di Cicco et al. 2017).

While HSMI has been formally diagnosed only in Atlantic salmon, other salmonids and some marine fish are also susceptible to infection by PRV (Wiik-Nielsen et al. 2012; Morton et al 2017; Purcell et al. 2018). In the last few years, there has been a surge of PRV-related diseases reported in Pacific salmon in Norway, Chile, Japan, and Canada (Olsen et al. 2015; Godoy et al. 2016: Takano et al. 2016: Miller et al. 2017). In addition, outbreaks of a disease designated as 'HSMI-like' were reported in farmed Rainbow trout (*Oncorhynchus mykiss*) across five hatcheries and two marine farms in Norway, and a PRV strain (variously termed $PRV_{om}$ or PRV-II, most recently classified as PRV-3) was sequenced from affected fish (Olsen et al. 2015; Gjevre et al. 2016; Hauge et al. 2017). Challenge studies using tissue homogenates from affected Rainbow trout resulted in the development of the pathological lesions, but as with HSMI, not the clinical signs (Hauge et al. 2017). However, while inflammatory lesions in the heart and skeletal muscle were present, heart inflammation was more concentrated in the spongy layer. In field outbreaks, some clinical signs were similar to HSMI, e.g. anorexia, lethargy, and ascites. Other features of the disease were not common in fish with HSMI, notably, anemia, jaundice (yellowish liver), hemosiderosis in the spleen, and hemorrhaging that are all consistent with lysis of red blood cells (RBCs). The same strain of PRV, and possibly PRV-1, has also been implicated in outbreaks in farmed Chilean Coho salmon (*Oncorhynchus kisutch*), in which major hepatic necrosis and erythrophagocytosis in the kidney and spleen were prominent pathological features commonly reported (Godoy et al. 2016).

Farmed Japanese Coho salmon have also been afflicted by a disease caused by PRV-2, a strain that is distinct based on nucleotide sequence from PRV-3 (i.e. the Norwegian PRV-II). The disease associated with PRV-2 was named erythrocytic inclusion body syndrome (EIBS) (Michak et al. 1992) where the inclusions are due to viral factories in the RBCs, but is highly similar in clinical and pathological signs to HSMI-like disease (Takano et al. 2016). Severe anemia and jaundice were hypothesized to be caused by excess bilirubin in the liver and a mechanism involving viral mediated lysis of RBCs was suspected (Sakai et al. 1994).

In British Columbia, Canada, the dominant farmed species is Atlantic salmon, while Chinook and Coho salmon, both endemic species, currently make up approximately 3% of farmed biomass. A disease characterized by jaundice and anemia, also called jaundice syndrome, has caused low level mortality of farmed Chinook salmon for several years (Garver et al. 2016), and as observed in Japan, Chile, and Norway, the disease is associated with PRV (Miller et al. 2017). In BC, however, only a single strain of PRV has been observed, PRV-1. This is the same strain that causes HSMI in Atlantic salmon.

As a cause and effect relationship has been demonstrated only between Strain PRV-1 and HSMI in Atlantic salmon and for Strain PRV-2 and EIBS in Japanese Coho, this study set out to resolve whether Strain PRV-1 is likely to play a causative role in the development of jaundice/anemia in BC Chinook salmon. This question is relevant to understanding the risk associated with PRV transmission from farmed to wild salmon in BC, a topic of interest to the public, government regulators, and industry.

# Methods

## Farm Samples

Farmed Chinook and Atlantic salmon were provided by the Fisheries and Ocean Canada (DFO) regulatory farm audit program. This program conducts randomized farm audits across the BC salmon aquaculture industry (Fig. S1) to assess the role of infectious disease in background losses to production populations and to ensure reporting compliance with OIE (World Organization of Animal Health) listed diseases. In each farm audit, one to nine fresh silver (recently dead) fish were provided from collections undertaken in 2011-2013, with 210 Chinook and 672 Atlantic salmon samples incorporated into our analysis. At the time of collection, clinical and environmental data and gross lesions were recorded, and tissue samples were taken for histopathology, bacterial and viral culture, as well as molecular analysis. Histopathological assessments of audit salmon were provided by Hugh W. Ferguson and veterinary diagnostics were provided by Ian Keith (DFO Aquaculture Management Division, Pacific Region).

The name “jaundice/anemia” assigned to the condition observed in BC Chinook salmon derives from the most striking clinical sign (yellowing of the abdomen and under the eye) occurring during the development of the disease, with anemia preceding the appearance of jaundice. Chinook salmon were classified as “jaundice” if the veterinary diagnostic comments indicated “Jaundice syndrome” or “jaundice - no agent”, and/or the gross lesions indicated “yellow fluid” or “yellow bile” or “yellow bile-like fluid” noted in the peritoneal cavity or on the pyloric caeca and/or liver. As well, two fish with pathological notes containing either “(inflammatory) lesions in heart or spongy layer only – CMS-like” or “acute renal tubular necrosis, suggestive of viral infection” were classified as “jaundice”, as these are also features of jaundice/anemia. Fish were classified as “anemia” if clinical data included the terms “anemic” or “pale liver” or “pale gills”.

While the international standard for diagnosing HSMI (Biering and Garseth 2012) is, at a farm-level, based on the presence of inflammatory lesions in heart and skeletal muscle, and not pancreas (principally to differentiate HSMI from other virally-mediated heart diseases in Norway), the audit sampling program did not include samples of skeletal muscle or pancreas tissue until late 2013. However, in a recent publication diagnosing HSMI for the first time on a BC salmon farm (Di Cicco et al. 2017), our group reported that the viral agents responsible for the other two heart diseases, cardiomyopathy syndrome (CMS) and pancreas disease (PD), were not detected in a surveillance of over 10,000 salmon in British Columbia. Therefore HSMI can be recognized in BC simply on the basis of virally-mediated myocarditis lesions in the heart. Hence, Atlantic salmon in this study were diagnosed with "HSMI" if the histopathology notes indicated "CMS/HSMI-like", or "CMS-like myocarditis", or "(inflammatory) heart lesions look to be viral", or "severe viral-type myocarditis/pericarditis", or "(inflammatory) heart lesions affecting compact layer".

A longitudinal collection of Atlantic salmon from a single farm outbreak of HSMI (Di Cicco et al. 2017) was also utilized in our analyses, providing a means to finely study the disease development pathway of HSMI in a single population. These collections included sampling of live and moribund/recently dead fish from a single farm in the Discovery Islands over the 11 month course of the disease outbreak. The pathological investigation showing the development of inflammatory lesions in both heart and skeletal muscle tissues within these sampled fish over time is already described (Di Cicco et al. 2017), and we used the classification of fish along the disease development pathway outlined in that study to identify individuals upon which to attempt the localization of PRV through *in situ* hybridization (ISH) analysis.

## Molecular Analyses

Samples for molecular analysis were collected through dissections on the farms, where live-sampled fish were killed by industry personnel by stunning (a blow to the head), as is the industry standard for food fish. For the longitudinal sampling, brain, liver, gill, kidney, and heart were sampled and preserved separately in RNA-later, held at $4^{o}$C for 24 hours before being transferred into a $-80^{o}$C freezer until nucleic acid extraction. Audit tissue samples also included spleen and pyloric caeca, with skeletal muscle and pancreas added to a small portion of the samples in 2013; audit tissue samples were kept on ice until reaching the laboratory by the end of the day and then frozen at $-80^{o}$C in a single sterile tube. DNA and RNA were extracted from combined Tri-reagent™ homogenates of all collected tissues for the audit samples, and solely the heart tissue for the HSMI longitudinal study, using methods described previously (Miller et al. 2016).

We employed the Fluidigm BioMark™ HD microfluidics quantitative PCR platform to assess the presence and load of 45 salmon infectious agents, including viruses, bacteria, fungi and protozoan parasites for Atlantic salmon in the longitudinal study (Di Cicco et al. 2017) and for farm audit salmon (Miller et al. 2017). Viruses with TaqMan assays included in this analysis: PRV, *Atlantic salmon paramyxovirus* (ASPV), *Infectious hematopoietic necrosis virus* (IHNV), *Infectious salmon anemia virus* (ISAV), *Pacific salmon parvovirus* (PSPV), *Piscine myocarditis virus* (PMCV), *Salmonid alphavirus* (SAV), *Salmonid herpesvirus* (OMV), *Encephalopathy and retinopathy virus* (VER), *Erythrocytic necrosis virus* (ENV), and *Viral hemorrhagic septicemia virus* (VHSV). Details on the analytical performance of these and other infectious agent assays on the platform were evaluated previously (Miller et al. 2016), in which the sensitivity (limit of detection), repeatability, and specificity of each assay are shown. To limit false positive amplification, all laboratory processing steps were carried out in dedicated workspaces, most in UV hoods, separating pre- and post-amplification procedures into different laboratories and adding artificial construct positive controls last, as described in Miller et al. (2016).

A panel of host biomarkers (genes) that when co-expressed identifies fish in a viral disease state (VDD) was developed and validated (Miller et al. 2017), where their application to the salmon audit samples using the Fluidigm BioMark platform was also described. The original panel evaluated by Miller et al. (2017) included TaqMan assays to 51 biomarkers, 20 of which had good efficiencies and provided viral disease resolution in Chinook salmon, and 22 in Atlantic salmon (Table S1). However, in the 2017 paper, we showed that as few as 11 biomarkers were required to differentiate fish merely carrying a virus from those experiencing cellular-level expression of a viral disease (and sensitive to the early phase of disease, when histopathological changes are not yet visible), which was the application of interest for this study. Moreover, we showed that the VDD signature was predictive of viral disease regardless of the tissue assessed, and as such, worked equally well on the multi-tissue RNA samples from the audit program as opposed to single tissue (heart) analysis.

We applied the VDD panel to the audit fish in order to resolve early disease development associated with PRV in both Atlantic and Chinook salmon, and to differentiate fish that were simply carriers of the virus from those that were actually in a viral disease state. The panel was applied to all of the fish in our collections, hence also included fish for which PRV was not detected, and could potentially contain other viral-mediated diseases. Herein, we restrict most of our presentation, and all of our ISH analyses (except for controls), to fish that contained PRV at loads $>10^3$ copies per µg nucleic acids (with one exception; most $>10^4$ copies per µg), although we calculate prevalence of a viral disease state in sampled farmed salmon.

Host biomarkers in the VDD panel were normalized to three reference genes derived from previous microarray studies (Table S1), and relative gene expression was assessed using the 2-ΔΔCt method (Livak and Schmittgen 2001). A pooled sample was used as the relative control. Classification of fish in a viral disease state was conducted through principle components analysis (PCA) of the expression data from 20 assays in Chinook and 22 assays in Atlantic salmon using the sample scores across PC1, as described by Miller et al. (2017). The goal of the VDD classification was to identify PRV-positive fish that may be in an early viral disease state, but not specifically diagnosed through pathology or clinical signs with a PRV-related disease. Hence, the edge of the VDD signature was determined by PC1 sample scores within the range of fish identified in a viral disease state (jaundice/anemia for Chinook and HSMI for Atlantic). A species-specific VDD threshold was chosen at the maximum value of Youden's J statistic for a ROC analysis of the PC1 sample scores, where viral disease state (identified/not identified) was used as the binary class for the ROC analysis. The Youden's J statistic represents the value of (sensitivity + specificity -1) for each point on the ROC curve and takes on values in [-1,1], with 1 characterizing a perfect test where sensitivity and specificity are both at 100. This approach optimizes the differentiating ability of the classifier when equal weight is given to sensitivity and specificity (Youden 1950).

## Sample selection for ISH

The goal of our ISH study was to track the movement of PRV within salmon tissues during early infection, development and peak states of disease development in Atlantic and Chinook salmon. To study HSMI in Atlantic salmon, we selected twenty four samples collected from the longitudinal study that diagnosed HSMI in British Columbia (Di Cicco et al. 2017) which had already been analyzed through histopathology and scored for heart and skeletal muscle lesions consistent with HSMI (see score system in Di Cicco et al. 2017). Following the indications of the previous analysis, we identified three stages of disease development, and conducted ISH analysis on six fish per stage. The stages were classified as:

1. Fish negative to PRV, and showing no histopathological lesions in any organ; negative control fish
2. Fish positive to PRV (>$10^3$ copies per µg nucleic acids), and showing no histopathological lesion in any organ (heart score = 0); "initial" stage of infection
3. Fish positive to PRV (>$10^3$copies per µg with one exception), showing mild degree lesions in the heart consistent with HSMI, but primarily localized in the epicardium and the compact layer of the myocardium (heart score = 1.5 to 3); "developing" stage of infection
4. Fish positive to PRV (>$10^3$-copies per µg), showing severe lesions in the whole heart (panmyocarditis) diagnostic of HSMI (heart score ≥4); "HSMI"

The samples chosen for ISH analysis consisted primarily of live, apparently healthy/asymptomatic fish for the first three groups, and moribund/freshly dead fish for the HSMI group.

Samples potentially representing different stages of development of jaundice/anemia in Chinook salmon were chosen on the basis of 1) PRV load (>$10^4$ copies per µg nucleic acids, with one exception), 2) VDD classification, and 3) whether fish had been diagnosed with jaundice/anemia. From these data, we selected three groups of samples:

1. Fish positive for PRV (>$10^3$ copies per µg), but VDD negative ; “PRV+ only” (three samples)
2. Fish positive for PRV (>$10^3$ copies per µg) and classified as in a VDD state; “PRV+/VDD+” (five samples)
3. Fish positive for PRV (>$10^3$ copies per µg), in a VDD state and diagnosed with jaundice (most also with anemia); “jaundice/anemia” (nine samples)

All audit Chinook samples were moribund/recently dead. No fish in group 1 and 2 showed signs of jaundice and only one fish in group 2 showed signs of anemia (but no jaundice).

## Histopathology

Preparation of the tissues for histopathological evaluation has been described previously (Di Cicco et al. 2017). Briefly, all tissues fixed in 10% neutral buffered formalin were dehydrated through an ascending gradient of alcohol solutions, embedded in paraffin wax, cut at 3.5µm thickness and stained with routine hematoxylin and eosin (H&E) for morphological evaluation by light microscope.

For both Atlantic and Chinook salmon, histology slides contained a section of heart (atrium and ventricle), liver, anterior and posterior kidney, spleen, intestine (often both small intestine and colon), skeletal muscle, brain and gills. All samples were read and scored by two pathologists (HWF and ED), with exception of group 1, 2 and 3 of the HSMI study which were read principally by one pathologist (ED). The scoring system for heart and skeletal muscle lesions were as described (Di Cicco et al. 2017), and followed Finstad et al. (2012). For the other organs, we utilized the scoring system used for regulatory purposes in the audit program. The severity of lesions was classified as 0 (no lesion), 1 (mild lesion), 2 (moderate lesion) and 3 (severe lesion). Consecutive slides from the same fish were cut for ISH to allow comparison between the presence of histopathological lesions and the localization of PRV.

Statistical evaluation of histopathology scores was undertaken to identify lesions showing the strongest inter-species differences. Principle components analysis was applied to the lesion scores to identify the main trajectories in the data and visualize whether Atlantic and Chinook salmon clustered separately. A PCA biplot was used for visualization to project the lesions onto the same graph as the samples (Ringnér 2008; Gabriel 1971). As well, lesion scores were treated as binary variables (zero score vs positive score) and contrasted between species for each of the three groups with PRV detections. Barnard’s (exact) test was used to identify lesions that were significantly different between species (Barnard 1945). The analysis was implemented in R version 3.3.2 based on the ‘Exact’ package (version 1.7).

## *In Situ* Hybridization (ISH)

RNA-ISH was performed using BASEscope (RED) (Advanced Cell Diagnostics, Newark, CA) according to the manufacturer's instructions. Briefly, consecutive sections of all Atlantic and Chinook salmon samples utilized for the histopathological analysis were dewaxed by incubating for 60 min at 60°C and endogenous peroxidases were quenched with hydrogen peroxide for 10 min at room temperature. Slides were then boiled for 30 min in RNAscope Target Retrieval Reagents and incubated for 30 min in RNAscope Protease IV reagent prior to hybridization. The slides underwent hybridization with a BASEscope probe against a portion of PRV-1 genome segment L1 (that codes for PRV core shell ; Advanced Cell Diagnostics, catalog #705151), in order to detect PRV in the tissues. A BASEscope probe against the bacterial gene dapB (#701021) was used as negative control to confirm absence of background and/or non-specific cross-reactivity of the assay. Concurrently, PRV/HSMI positive (n = 2) and negative (n = 1) heart samples from Atlantic salmon collected during an experimental challenge for HSMI performed in Norway (Finstad et al. 2014) were used as positive and negative controls for PRV-1 presence and localization. Signal amplification was performed according to the manufacturer's instructions, followed by counterstaining with Gill's hematoxylin and visualization by bright field microscopy.

## Immunohistochemistry with antibody to hemoglobin

To evaluate the presence of hemoglobin in the damaged tissues of Chinook salmon, an antibody raised against hemoglobin (rabbit polyclonal Anti-Hemoglobin subunit beta antibody, ab202399, Abcam Inc., Toronto, Canada) was utilized to perform immunohistochemistry on a subset of samples of Chinook salmon (affected by various degrees of renal tubular necrosis – n=3 fish classified as "PRV+/VDD+" and n=3 samples from fish diagnosed with jaundice/anemia ) and fish presenting no lesions in the kidney (n=2 fish classified as "PRV+ only" ). The protocol was previously described (Di Cicco et al. 2017). Briefly, 3.5 µm thick paraffin wax embedded sections were mounted on Superfrost Plus glass slides (Thermo Fisher Scientific, Portsmouth, USA). These were heated at 60 °C for 20 min, dewaxed in xylene and rehydrated through graded alcohols. Antigen retrieval was performed by autoclave treatment (121 °C for 10 min) in citrate buffer (0.1 M, pH 6.0). The sections were successively treated with 3% hydrogen peroxide for 1h to block endogenous peroxidase activity. Non-specific binding sites were blocked by normal goat serum (Vector Laboratories, Burlingame, CA, USA) diluted 1:10 in 1% bovine serum albumin (Vector Laboratories, Burlingame, CA, USA) in TBS [pH 7.6, 0.05 M Tris/HCl, 0.15 M NaCl] for 1h. The same diluent solution was used for primary antibodies (1:100) and then incubated in a humidity chamber at 4 °C overnight. A Vectastain ABC-peroxidase kit (Vector Laboratories, Burlingame, CA, USA) was used for detection of bound antibody according to the manufacturer's instructions, and ImmPACT NovaRED (Vector Laboratories, Burlingame, CA, USA) was utilized as substrate. Finally, the sections were counterstained with Harris' hematoxylin and mounted (Permount). All incubations, except with the primary antibodies, were carried out at room temperature in a humidity chamber. Two PRV negative samples (including all internal organs and not showing histopathological signs of hemolysis and/or anemia) were used as negative control. Primary and secondary antibody controls were performed by replacing the antibody with the respective diluent alone.

## Whole genome sequencing of PRV

RNA viruses exist as groups of closely related genotypes, known as a viral quasispecies. These can exist as several master variants, each with their own cloud of mutants (Lauring and Andino 2010; Mordecai et al. 2016). Dhamotharan et al. (2018) recently showed that there are currently 3 strains of PRV, known as PRV-1, 2 and 3, which they described as variants of PRV. However, in order to accurately describe the variation found within these three groups, the nomenclature needs to be refined. In this manuscript we refer to PRV-1, 2 and 3 as strains (i.e. master variants). Finally, we refer to each unique consensus sequence (itself made up of individual sequencing reads from different individual virions) as a sequence variant.

While we know from previous studies that currently only Strain PRV-1 has been observed in British Columbia, we conducted whole viral genome sequencing in individuals across farm audits with fish diagnosed with HSMI (Atlantic salmon; 6 farms) or jaundice/anemia (Chinook salmon; 4 farms). For Chinook salmon, we also sequenced three individuals from farms with nearly all fish collected classified as PRV+/VDD+. This information is critical to determine if there were any consistent differences in the sequence of the virus associated with each disease, and consequently indicate the potential risk of disease transmission from farmed Atlantic to wild Pacific salmon.

For each farm audit, we sequenced the fish carrying the highest load of PRV (with one exception). We made every attempt to represent disease occurrence broadly distributed over the farming region in British Columbia, as dictated by the 2011 to 2013 audit samples provided to us. For Atlantic salmon, farms were located along the inside passage of Vancouver Island (Discovery Islands [Regulatory area A3.2]) and Broughton Archipelago [A3.3] and the Central Coast [A3.5] (Fig. S1). These audit collections spanned three seasons, represented three companies, included samples from 2011, 2012 and 2013, and included two collections from the same farm over successive seasons within the same year. For Chinook salmon, four farms were on the west coast of BC [P2], and three were located in the Discovery

Island/Broughton Archipelago region [P3] (see locations on Fig. S1). These audit collections spanned three seasons, represented four companies, included samples from 2011, 2012 and 2013, and included sampling from the same site and season collection over two separate years.

Because loads of PRV were very high for Chinook salmon (> $10^5$ copies/µg nucleic acids), we carried out RNA-sequencing with no enrichment, as described by Di Cicco et al. (2017). However, previous studies in our lab have suggested that full genome sequences were not guaranteed at loads < $10^5$ copies/µg nucleic acids, which was the case for most of the Atlantic salmon samples. Hence, we included a Sure-Select™ (Agilent, Santa Clara, CA) enrichment step for viral content (our Sure Select panel includes oligos across the full genomes of PRV-1 as well as many other salmonid viruses assessed on our infectious agent monitoring platform).

The Atlantic salmon samples were prepared using the SureSelect[XT] RNA Direct NGS target workflow (Agilent, Santa Clara, CA). This included preparing the RNAseq library with the SureSelect Strand-Specific RNA library Prep kit (Agilent, Santa Clara, CA) according to manufacturer's instructions. The adaptor ligated samples were purified with the Agencourt AMPure XP system (Beckman Coulter, Brea, CA). HS DNA chips were run on the Agilent 2100 Bioanalyzer (Agilent, Santa Clara, CA) to determine the final library size and the Qubit dsDNA HS kit (Invitrogen, Carlsbad, CA) was used to determine the concentration. Vacuum centrifugation (30°C/20') was necessary to obtain 200ng of the adapted libraries in the appropriate volume for hybridization. Hybridization of the adapted cDNA library with the viral SureSelect bait capture library (Agilent, Santa Clara, CA ) was performed at 65°C for 24hrs, according to manufacturer's instructions. The cDNA library/capture library hybrids were captured on streptavidin magnetic beads and purified with the Agencourt AMPure XP system (Beckman Coulter, Brea, CA). Index tags were added to the post captured libraries through 15 rounds of amplification and purified using the Agencourt AMPure XP system (Beckman Coulter, Brea, CA). HS DNA chips were run on the Agilent 2100 Bioanalyzer (Agilent, Santa Clara, CA) to determine the final library size and the concentration was determined using the Qubit dsDNA HS kit (Invitrogen, Carlsbad, CA). Sample libraries were normalized to 4nM, and denatured and diluted to obtain a final library of 20pM. The Atlantic salmon enriched RNAseq libraries were processed on 1 paired end v2 300bp kit on the Illumina MiSeq System (Illumina, San Diego, CA), which included a 10% phiX spike in.

For the Chinook salmon RNA-seq, ribosomal removal was performed using the Epicentre ScriptSeq Complete Gold Kit (Epidemiology) (Illumina, San Diego, CA) according to manufacturer's instructions. A Zymo RNA Clean and Concentrate-5 kit (Zymo Research, Irvine, CA) was applied to purify the rRNA depleted total RNA, according to manufacturer's instructions, and quantified using the Qubit RNA HS kit (Invitrogen, Carlsbad, CA). ScriptSeq Index reverse primers were added to the cDNA during the final amplification step, which involved 14 cycles. The Agencourt AMPure XP system (Beckman Coulter, Brea, CA) was applied to purify the 3'-terminal tagged cDNA and final amplified library. HS DNA chips were run on the Agilent 2100 Bioanalyzer (Agilent, Santa Clara, CA) to determine the final library size and the concentration was determined using the Qubit dsDNA HS kit (Invitrogen, Carlsbad, CA). Sample libraries were normalized to 4nM, and denatured and diluted to obtain a final library of 20pM. The Chinook RNA-seq libraries were run over 2 paired end v3 600bp reagent kits on the Illumina MiSeq System (Illumina, San Diego, CA), which included a 2% phiX spike in.

## PRV Genome Sequence Analysis

The quality of the raw reads was checked using FASTQC (https://www.bioinformatics.babraham.ac.uk/projects/fastqc/). Low quality reads or regions of reads and adapter sequences were removed using Trimmomatic (Bolger et al. 2014). The remaining reads were aligned to the PRV genome segments of the Norwegian isolate Salmo/GP-2010/NOR (Palacios et al. 2010) using the BWA-mem aligner with default parameters. A consensus sequence of the aligned reads was created within Geneious (V10.1.3) using a threshold of 95%. The consensus sequences were then aligned using Muscle (within Geneious), and trimmed to the shortest resulting contiguous sequence. Trees were constructed using the MrBayes plug in within Geneious.

# Results

## VDD classification in audit salmon

### *Atlantic salmon*

In farm audit data, HSMI was diagnosed across multiple individuals on four farms, with one farm showing evidence of HSMI in two year-classes. In all, 17 fish were specifically diagnosed with HSMI. In 11 of the HSMI fish, copy number of PRV in the mixed tissue sample was between $10^4$-$10^5$ per µg nucleic acids, but in six fish, it was $<10^4$. In all but two cases where PRV values were $<10^4$ but fish were diagnosed with HSMI, there were other fish on the same farm with HSMI with higher copy numbers of PRV. In one case, the fish was characterized with "Mild CMS-like lesions" and all other fish on the farm carried higher copy numbers of PRV. In the other case, a single fish was characterized as "Heart lesions are viral-like (CMS/HSMI)", but no PRV detection was observed on the farm; we suspect this was not an HSMI fish.

Principle components analysis (PCA) was applied to differentiate fish in a viral disease state based on the application of the VDD biomarker panel. One outlier fish with a rotational value of -24 was removed from the analysis. 90% of fish with PRV loads > $10^5$ and all but one fish diagnosed with HSMI had rotational values <0 (Fig. 1a). We established a VDD threshold at the point where the maximum of Youden's J statistic was observed for a ROC analysis of the rotational values, with HSMI disease vs no disease as the two classes. This approach optimized the differentiating ability of a HSMI classifier with equal weight given to sensitivity and specificity. This resulted in a VDD threshold at a rotational value of -2.42. At this threshold, 89% of Atlantic salmon with PRV copy number per µg nucleic acids between $10^5$-$10^6$ (19 fish) and 49% of fish between $10^3$-$10^5$ (70 fish) classified as VDD+ (Fig. 1c). 15% of fish with PRV loads exceeding $10^3$ classifying as VDD+ were diagnosed with HSMI. In all, 31% of 670 Atlantic salmon audit fish with biomarker data available were classified as VDD+; 43% of VDD+ fish contained multi-tissue PRV copy numbers exceeding $10^3$, 25% > $10^4$, and 9% >$10^5$. Only 16% of fish *not* classifying as VDD+ carried PRV exceeding $10^3$ and 5% exceeding $10^4$ copies per µg nucleic acids.

At a farm-level, the PRV-VDD classification revealed an additional 15 farms with multiple fish classified as PRV-VDD and 2 farm audits that had a single fish diagnosed with HSMI had at least one additional fish classifying as PRV-VDD; the outlier HSMI fish not in a VDD state was on one of these farms. In total, 20 of the 165 (12.1%) Atlantic salmon farm audits (excluding 38 audits with only a single fish sampled) carried farm-level evidence of HSMI or PRV-VDD, or both.

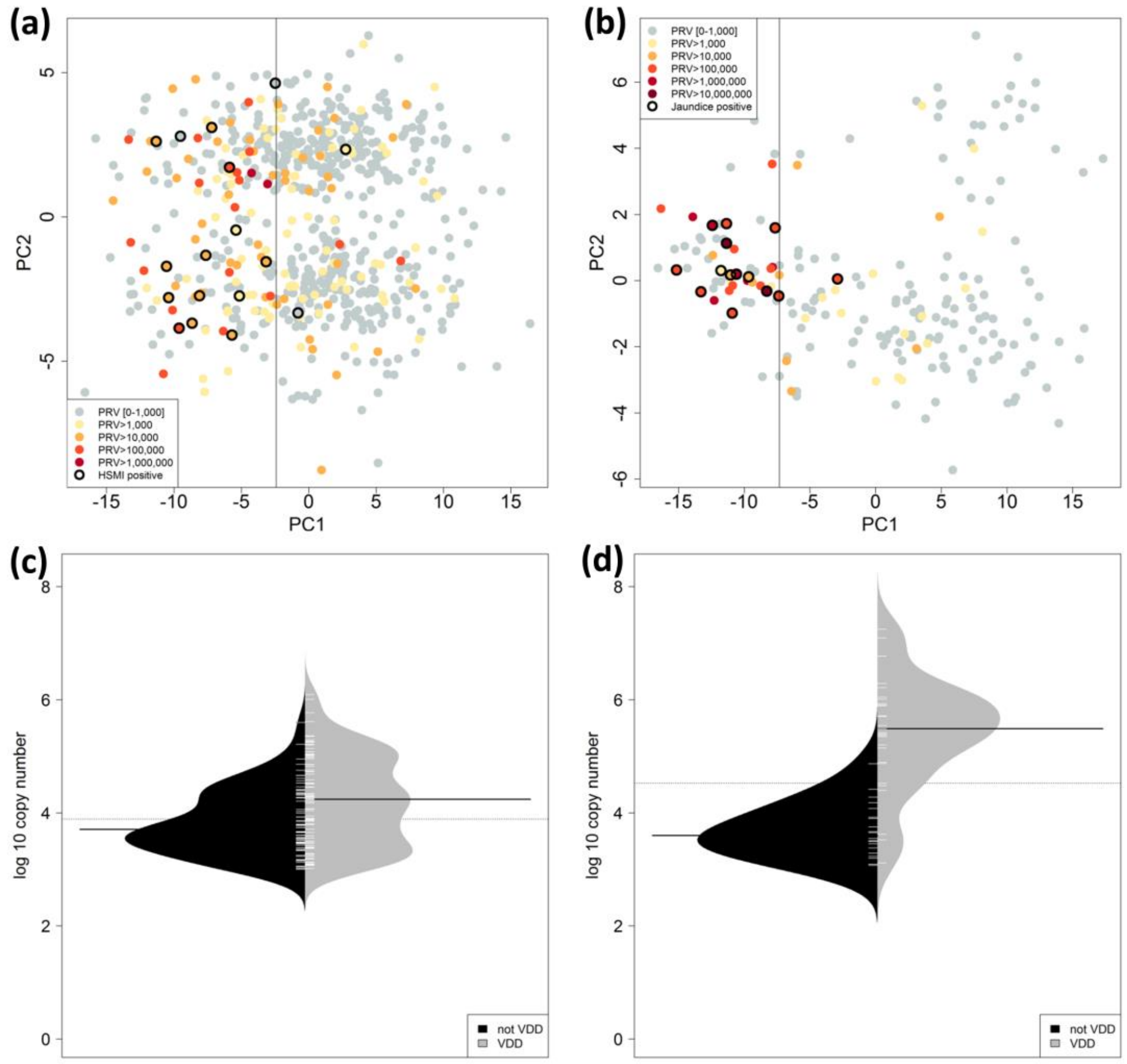

**Fig. 1**. Principle components analysis (a,b), and bean plots of the distribution of PRV productivity (log10 copy number) for VDD/no VDD groups (c,d) for viral disease development panels applied to farmed Atlantic salmon on the left (a,c) and Chinook salmon on the right (b,d). Principal component analysis show samples colored according to load of PRV with disease-diagnosed fish circled in black. The vertical line indicates the VDD threshold as determined by Youden's J statistic. Polygons in the bean plots represent non-parametric density estimates, white bars show total samples corresponding to log10 copy numbers, solid black bars represent the median log10 copy number in the middle plots, and mean log10 copy numbers in the bottom plots. The dotted line in the middle plots marks the overall median log10 copy number.

### *Chinook salmon*

There was evidence of jaundice/anemia across most, if not all collected samples in two of the Chinook salmon farm audits, with two additional farms containing single fish diagnosed with jaundice/anemia. In all, 13 Chinook salmon contained the clinical signs of jaundice/anemia and one fish only had clinical signs of anemia, but was from one of the farms with jaundice/anemia fish, and hence was reclassified as a jaundice/anemia fish (likely at an early stage of disease development); PRV copy number exceeded $10^4$ copies per µg nucleic acids in all but one of these fish.

All but one of the 14 Chinook salmon diagnosed with jaundice/anemia (and one with only anemia) clustered on the negative end of the PC1 distribution, with rotational values <-7.33; this was the VDD threshold determined the same way as for Atlantic salmon based on Youden's J statistic (Fig. 1b). There was a stronger threshold response in Chinook salmon to high PRV copy number, with 100% of fish with copy numbers exceeding $10^6$ (8 fish) and 93% from $10^5$-$10^6$ (14 fish) classifying as VDD+ (Fig. 1d). Only 29% of Chinook salmon with PRV copy number between $10^2$ and $10^3$ (28 fish) classified as VDD+. Over all Chinook salmon audit fish, 27% of fish with biomarker data classified as VDD+, with 46% of these associated with high loads of PRV (>$10^4$). 45% of PRV-VDD fish were diagnosed with jaundice/anemia.

At a farm-level, in both farms with singleton fish diagnosed with jaundice/anemia, additional fish sampled on the farm were classified as PRV-VDD. Three additional farms contained multiple fish classified as PRV-VDD. In total, 6 of 34 (17.6%) of farm audits with more than one fish sampled carried farm-level evidence of jaundice/anemia or PRV-VDD, or both.

## Infectious agent monitoring

### Atlantic salmon

Samples chosen for histopathology and ISH analysis of fish during an HSMI outbreak in Atlantic salmon (longitudinal study) were assessed for presence and load of 45 infectious agents in heart only. Only two agents other than PRV were present in some of the fish, both parasites, and neither correlated with developing heart lesions (presented by Di Cicco et al. (2017)). *Kudoa thyrsites* was detected in a portion of fish in all groups, with highest loads observed in two of the control (group 1) fish (Table 1). *Desmozoon lepeoptherii* (aka *Paranucleospora theridion*) was observed in all fish, with similar load distributions in each group.

### Chinook salmon

Agent distributions were more complex in the multi-tissue audit samples of Chinook salmon (Table 2). Recognizing that all of these fish were collected as moribund or recently dead, unlike the primarily live-sampled fish analyzed in the HSMI outbreak, they carried a larger range of agents, some with pathological changes suggesting mixed infections (Table 3). Single instances of rickettsiosis, bacterial kidney disease, vibriosis, aspecific branchitis and *Loma* infection were diagnosed and verified by the presence of the causative pathogens (at high load). Two of these individuals were in group 1 (PRV+/VDD-), while the remainder were in group 2 (PRV+/VDD+). In one instance, while rickettsiosis was a differential, no *P. salmonis* was observed (772). The only virus that was observed other than PRV was an occasional detection of the DNA virus ENV. Miller et al. (2017) showed that the VDD panel was not stimulated in response to the presence of ENV, and as it was designed only with transcriptome data for RNA viruses, we speculated that it would not identify a DNA viral disease state; however this will need to be confirmed by further analysis. Hence, while three individuals classified as jaundice/anemia contained high loads of ENV, ENV was unlikely to have contributed to their molecular classification. Moreover, t-tests to assess whether any infectious agent(s) other than PRV were statistically associated with jaundice/anemia revealed no significant results.

## Histopathology and ISH

The scoring of heart lesions diagnostic of HSMI in Atlantic salmon is presented in Table 1. The type and severity of histopathological lesions reported are shown across all tissues in Atlantic salmon during the course of an HSMI outbreak and Chinook salmon during the development of jaundice/anemia in Table 3. Note that while we were able to provide a quantitative score (0-3) for each pathological lesion, due to the large number of different cell types positively marked by ISH (i.e. infected by PRV) that appeared to be involved in the pathogenesis of the two diseases, and the different amount and size of positive dots observed in the cells (indicating a different quantity of viral particles, incomplete particles and viral transcripts in a determined area), we carried out only a qualitative assessment of PRV presence, localization and quantity, to describe where the virus was located during the different stages of development of the two diseases.

Below, we provide a description of the disease development process within each of these diseases and the localization of PRV through ISH.

## Atlantic salmon

Overall, in the longitudinal study, we did not observe any variation in the quality of PRV staining between live, moribund and recently dead individuals, suggesting that variation in tissue quality between species (Chinook were all moribund/recently dead) is not likely to impact our findings.

**Blood**: PRV was detected in RBCs from all fish with molecular detection of PRV, and none where PRV was not detected by qPCR. The marking was characterized by several dots of various sizes inside the cytoplasm of the cells (Fig. 2a). In some cases, the dots were so numerous they filled the entire cytoplasm of the RBCs. The marking in the RBCs was particularly evident in fish at the initial phase of infection, while those classified as "developing" or "HSMI" showed, overall, a slight decrease of marking.

**Heart**: No lesions were reported in the heart of fish that did not carry PRV or those in the initial phase of the infection with high loads of PRV (Fig. 3a). PRV marking was predominantly localized in several RBCs of the blood circulating inside the organ in the initial phase of infection (Fig. 3b), while rare, scattered dots were observed in few cardiomyocytes in two of six fish. All fish classified as "developing lesions" showed mild to moderate inflammation primarily involving the epicardium, but a few inflammatory foci were also present in the compact layer and, to a minor extent, in the spongy layer of the myocardium (Fig. 3c, 4a and 4c). In these samples, PRV marking was observed in a few endothelial cells from small vessels as well as nearby cardiomyocytes in both compact and spongy layers of the myocardium. These areas of PRV-infected cardiomyocytes were often associated with small aggregates of lymphocytes and macrophages, although several infected cardiomyocytes were also scattered in the tissue, with no associated inflammatory reaction (Fig. 3d, 4b and 4d). The marking of cardiomyocytes was characterized by several dots of various sizes distributed inside the cytoplasm of the cells. Fish diagnosed with HSMI had extensive inflammation in the heart, involving epicardium, both layers of the ventricular myocardium, and in some instances, the atrium (i.e. panmyocarditis) (Fig. 3e and 4e). The inflammation was morphologically characterized by the infiltration of lymphocytes and macrophages, and was associated with scattered endocardial hypertrophy and intercellular edema (between endocardial and myocardial cells, and in particular in the spongy myocardium, when severely inflamed) and focal areas of myocardial degeneration indicated by loss of striation and hypertrophic cardiomyocytes as signal of damage reparation. In HSMI fish, PRV was widely distributed in most of the cardiomyocytes of the compact layer, but several extended areas of the spongy myocardium were also heavily marked (Fig. 3f and 4f). At times, the dots in the cardiomyocytes were so numerous as to basically fill the whole cytoplasm of the cells. Severe, extensive inflammatory infiltrates were associated with the PRV-infected areas.

**Liver**: Fish not diagnosed with HSMI showed no apparent lesions in the liver, with the exception of rare, small, and scattered hepatocellular necrotic foci in fish classified as "developing" (Fig. 5a and 5c). PRV-marking was confined to RBCs in fish in the initial phase of the infection (Fig. 5b). In the "developing" phase, several hepatocytes were also infected with PRV (Fig. 5d), with marking dots of various sizes and number distributed inside the cytoplasm of the cells. All fish diagnosed with HSMI presented widespread areas of moderate hepatocellular necrosis, with the typical distribution pattern occurring in a "heart failure" condition (i.e. zonal necrosis primarily localized around the central veins, which represent the farthest areas of the liver parenchyma from oxygenated blood) (Fig. 5e); the number of positively marked hepatocytes was slightly reduced in HSMI fish, but still present in several cells (Fig. 5f).

**Kidney**: No necrotic lesions were reported in the kidney from any of the groups (Fig. 6a). However, fish classified in the "developing" phase of HSMI and fish diagnosed with HSMI presented with some cases (2/6 and 3/6, respectively) of mild interstitial hyperplasia, and an increased number of leucocytes in the blood vessels (Fig. 6c and 6e). PRV was detected in several RBCs in the kidney, but also in a few macrophages and melanomacrophages scattered through the hemopoietic tissue in fish in the initial phase of infection (Fig. 6b), although the number of macrophages positive to PRV marking was notably higher in two of the six fish. In fish classified as "developing", the number of macrophages and melanomacrophages positive for PRV increased considerably, particularly around the kidney tubules (Fig. 6d), and slightly decreased in fish with HSMI (Fig. 6f).

**Spleen**: No lesions were observed in the spleen of fish that were PRV negative or in the initial phase of infection (Fig. 7a). In the initial phase, PRV was present in several RBCs and in fewer reticular cells of the ellipsoids and around the sinusoids (Fig. 7b). In fish classified as "developing" and fish diagnosed with HSMI, mild white pulp hyperplasia was reported (2/6 and 3/6, respectively) (Fig. 7c). All fish with HSMI also had mild to severe ellipsoid necrosis, with loss of reticulin framework, and mild to moderate congestion in 4 of 6 fish (Fig. 7e). The number of PRV-infected RBCs increased substantially (highlighting the localization in the parenchyma of the sinusoids, which appeared full of RBCs) in fish classified as "developing" and was associated with an increase in PRV-positive macrophages around the sinusoids (Fig. 7d); this pattern was slightly decreased in fish with HSMI (Fig. 7f).

**Intestine**: No lesions in the intestine were observed in any of the four groups (Fig. S2a, 2c and 2e). Positively marked RBCs were observed in all fish with molecular detection of PRV (Fig. S2b), but several enterocytes were also marked in fish classified as "developing" (Fig. S2d). The number of PRV positive enterocytes was considerably lower in fish with HSMI (Fig. S2f), and the marking was characterized by a few small dots scattered in the cytoplasm of the enterocytes.

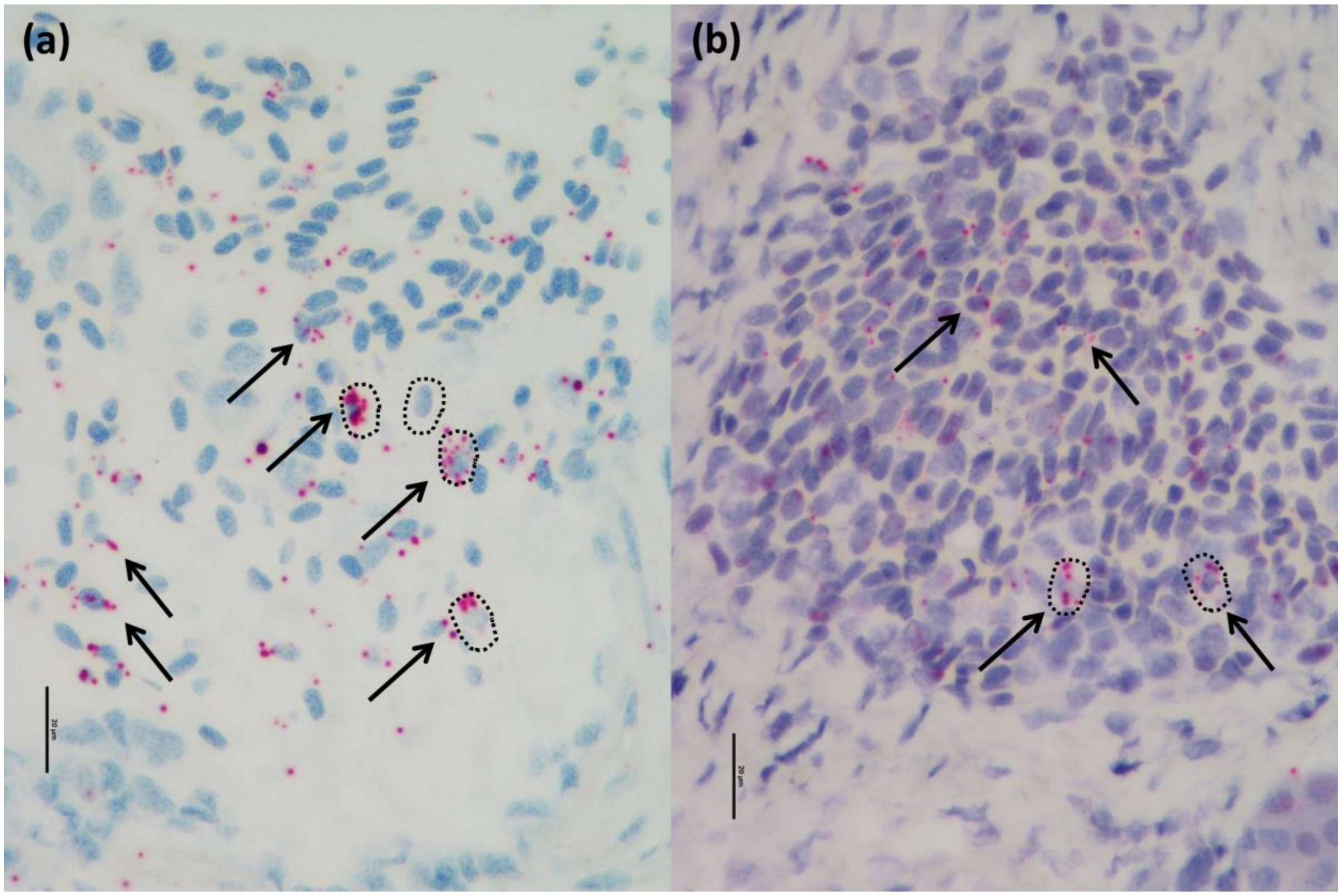

**Fig. 2.** a) Blood clot in Atlantic salmon with no evidence of lesions. Several PRV particles (red – arrows) are visible in numerous RBCs (some of which are defined by the dotted circles). ISH (scale bar 20µm). b) Blood clot in Chinook salmon with no evidence of VDD or lesions. Several PRV particles (red – arrows) are visible in numerous RBCs (some of which are defined by the dotted circles). ISH (scale bar 20µm).

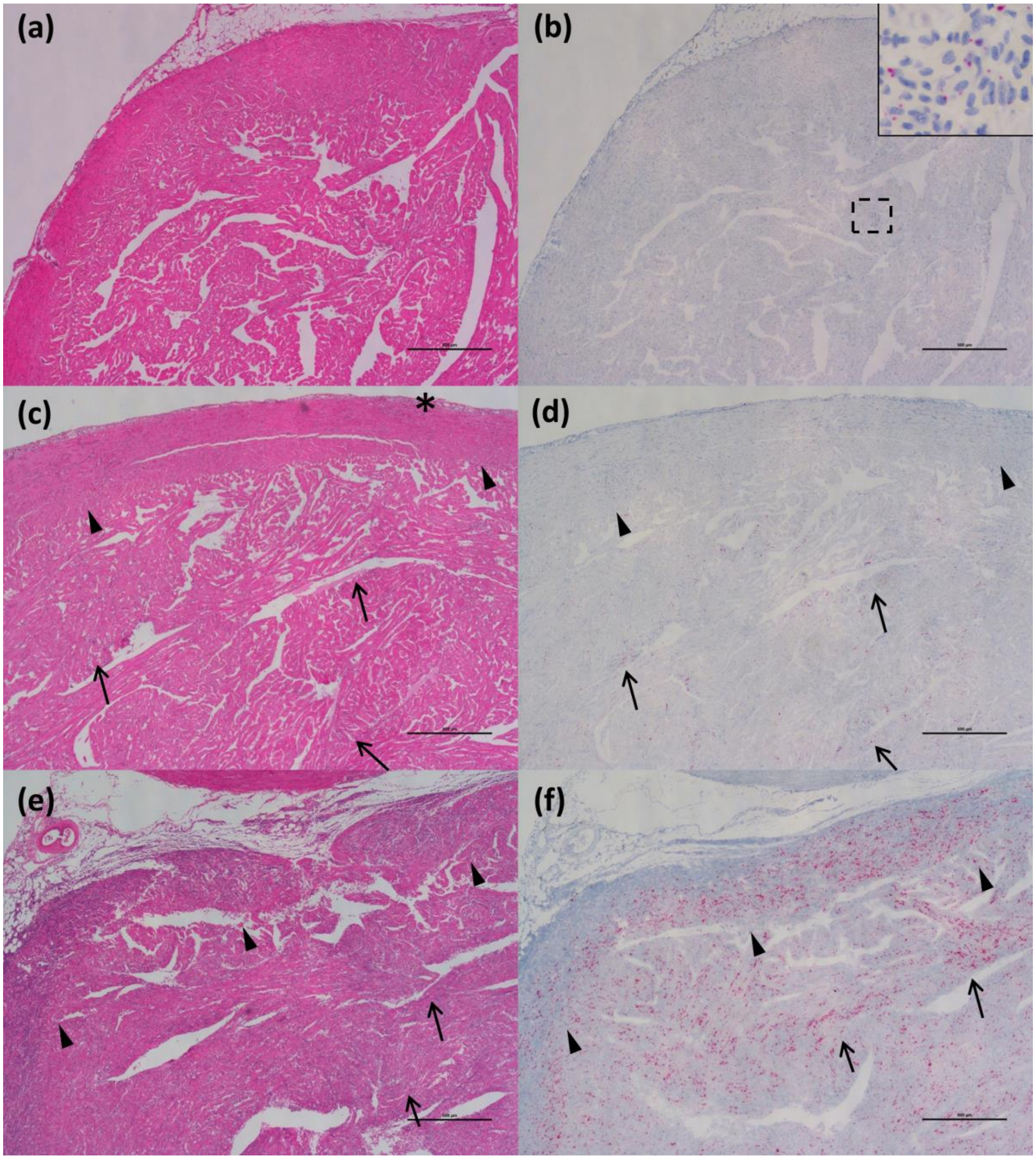

**Fig. 3.** Histopathology and ISH in Atlantic salmon heart. Matching fields of consecutive sections from samples at various stages of HSMI disease development are stained with H&E (a, c, e) and ISH for PRV-1 L1 segment (b, d, f). The scale bar in all figures is 500µm. a) Group 2, with no apparent lesions. b) PRV (red) was localized in the RBCs only (inset). c) Group 3, showing moderate epicarditis (star) and few inflammatory infiltrates in the compact layer (arrowheads) and spongy layer of the myocardium (arrows). d) Nodules of PRV positive cardiomyocytes (red) are localized in the compact layer (arrowheads) and in the spongy layer (arrows) of the myocardium, particularly around blood vessel and associated with inflammatory infiltrates, although several cardiomyocytes also show no nearby inflammation. e) Group 4, showing severe epicarditis and widespread endocarditis, involving virtually all the compact layer (arrowheads) and extensive areas of the spongy layer (arrows) of the myocardium. f) A large number PRV positive cardiomyocytes (red) were observed virtually across all the compact layer of the myocardium (arrowheads), although several of PRV positive cardiomyocytes were also observed in the spongy layer, often associated with inflammatory infiltrates.

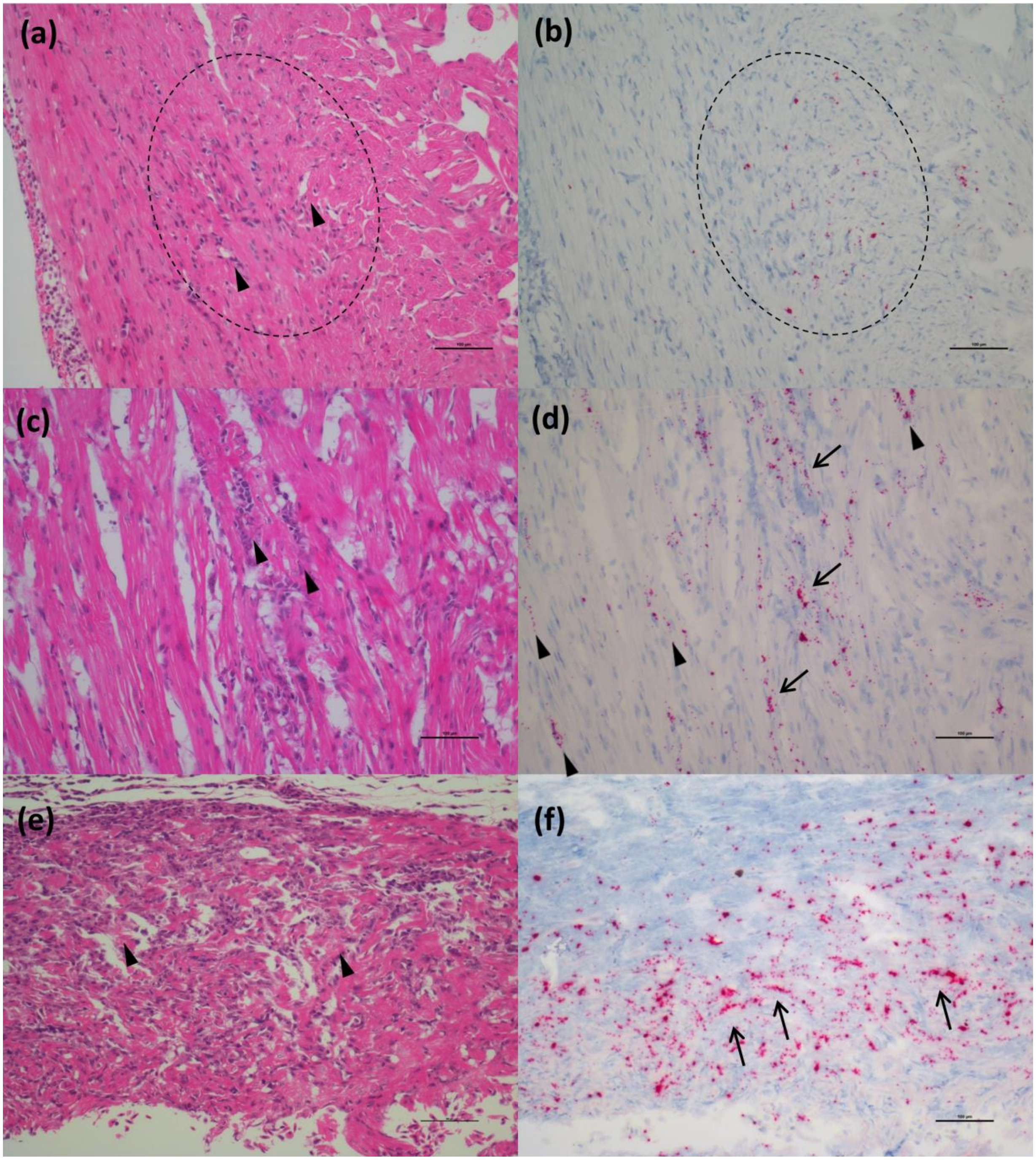

**Fig. 4.** Histopathology and ISH in Atlantic salmon heart. Matching fields of consecutive sections from samples at various stages of HSMI disease development are stained with H&E (a, c, e) and ISH for PRV-1 L1 segment (b, d, f). The scale bar in all figures is 100µm. a) Group 3, showing nodule of inflammatory infiltrate (circle) in the compact layer of the myocardium, in proximity to two small blood vessels (arrowheads). b) PRV (red) was localized in cardiomyocytes involved by the inflammatory infiltrate (circle). c) Group 3, showing inflammatory infiltrates in the spongy layer of the myocardium (arrowheads). d) PRV positive cardiomyocytes (red) were involved by the inflammatory infiltrates, although several cardiomyocytes also showed no nearby inflammation (arrowheads). e) Group 4, showing a severe, chronic endocarditis and myocardial degeneration (arrowheads) across the whole compact layer of the myocardium. f) A large number PRV positive cardiomyocytes (red) were observed virtually across the whole compact layer of the myocardium. Several cardiomyocytes are replete with PRV (arrows).

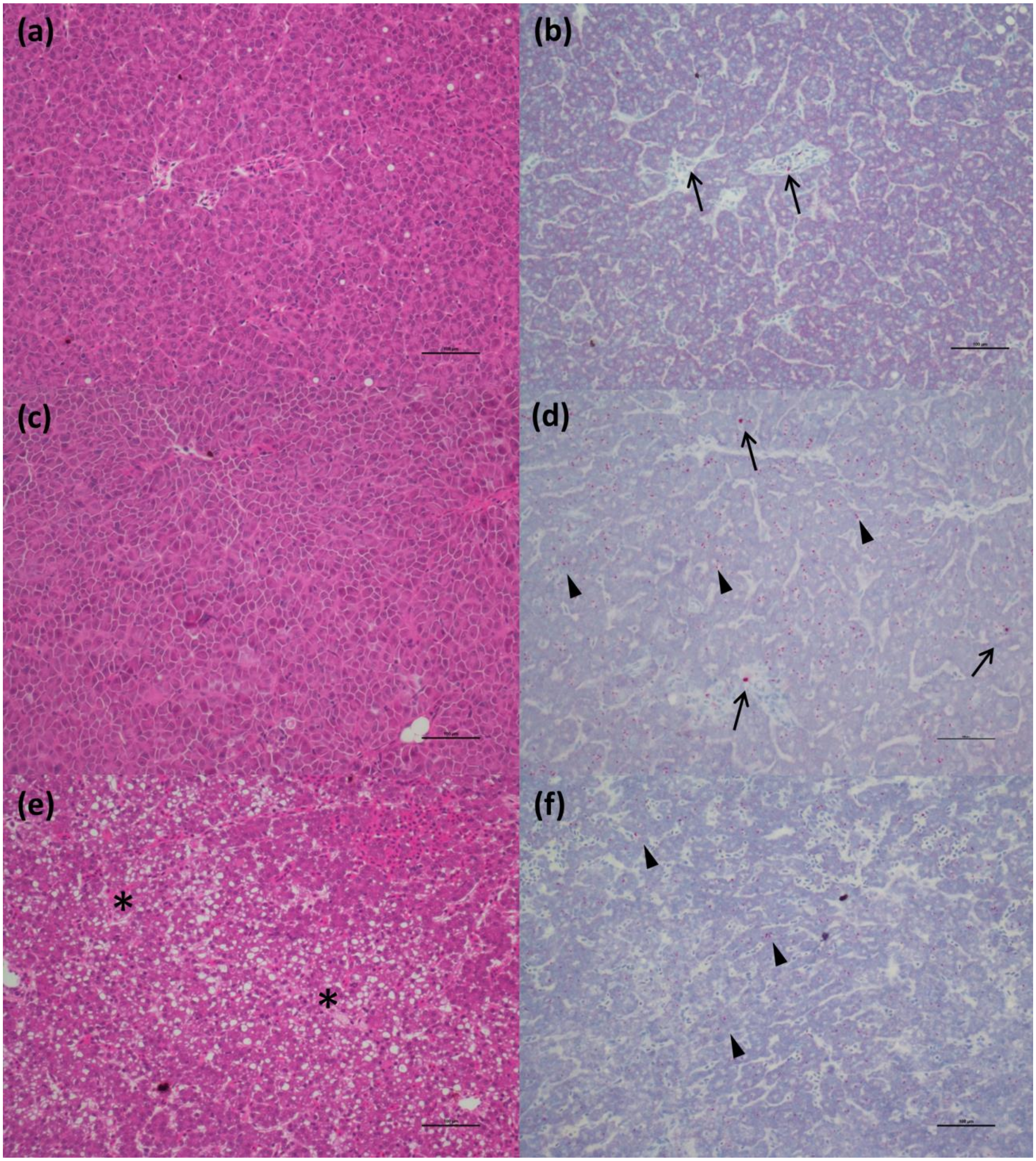

**Fig. 5.** Histopathology and ISH in Atlantic salmon liver. Matching fields of consecutive sections from samples at various stages of HSMI disease development are stained with H&E (a, c, e) and ISH for PRV-1 L1 segment (b, d, f). The scale bar in all figures is 100µm. a) Group 2, with no apparent lesions. b) PRV (red) was localized in the RBCs (arrows) only. c) Group 3, with no apparent lesions. d) PRV (red) was present in the RBCs (arrows), but was mostly localized in the hepatocytes' cytoplasm (arrowheads). e) Group 4, showing extensive areas of hepatic necrosis (stars), mostly presenting the typical "heart failure" pattern of distribution. f) PRV (red) was still localized mostly in the hepatocytes' cytoplasm (arrowheads).

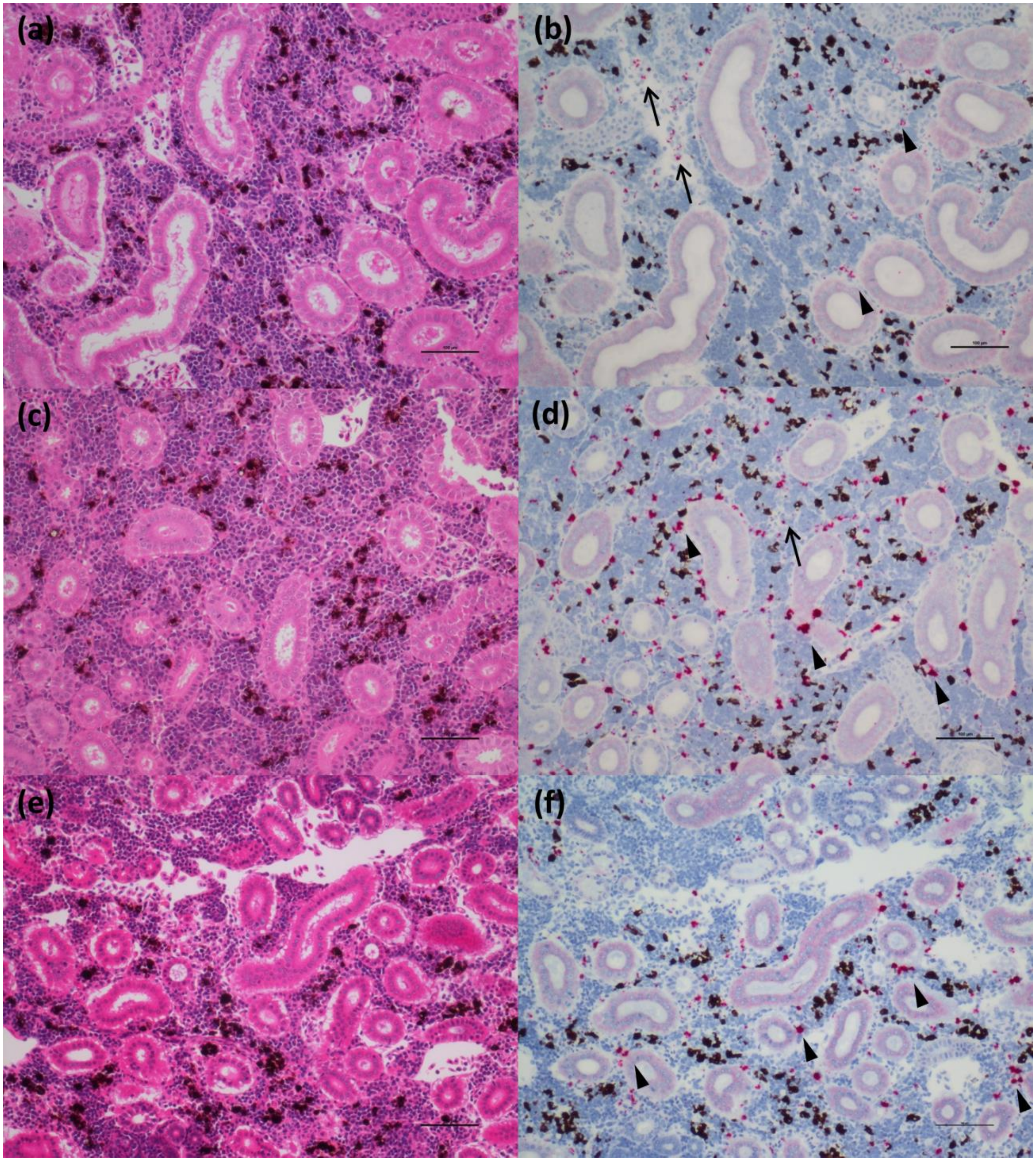

**Fig. 6.** Histopathology and ISH in Atlantic salmon posterior kidney. Matching fields of consecutive sections from samples at various stages of HSMI disease development are stained with H&E (a, c, e) and ISH for PRV-1 L1 segment (b, d, f). The scale bar in all figures is 100µm. a) Group 2, with no apparent lesions. b) PRV (red) was localized in the RBCs (arrows), but also a few macrophages (arrowheads). c) Group 3, showing mild interstitial hyperplasia. d) PRV (red) was present in the RBCs (arrow), but was mostly localized in the peritubular (renal portal) macrophages (arrowheads). e) Group 4, showing mild interstitial hyperplasia. f) PRV (red) was still localized mostly in the macrophages (arrowheads).

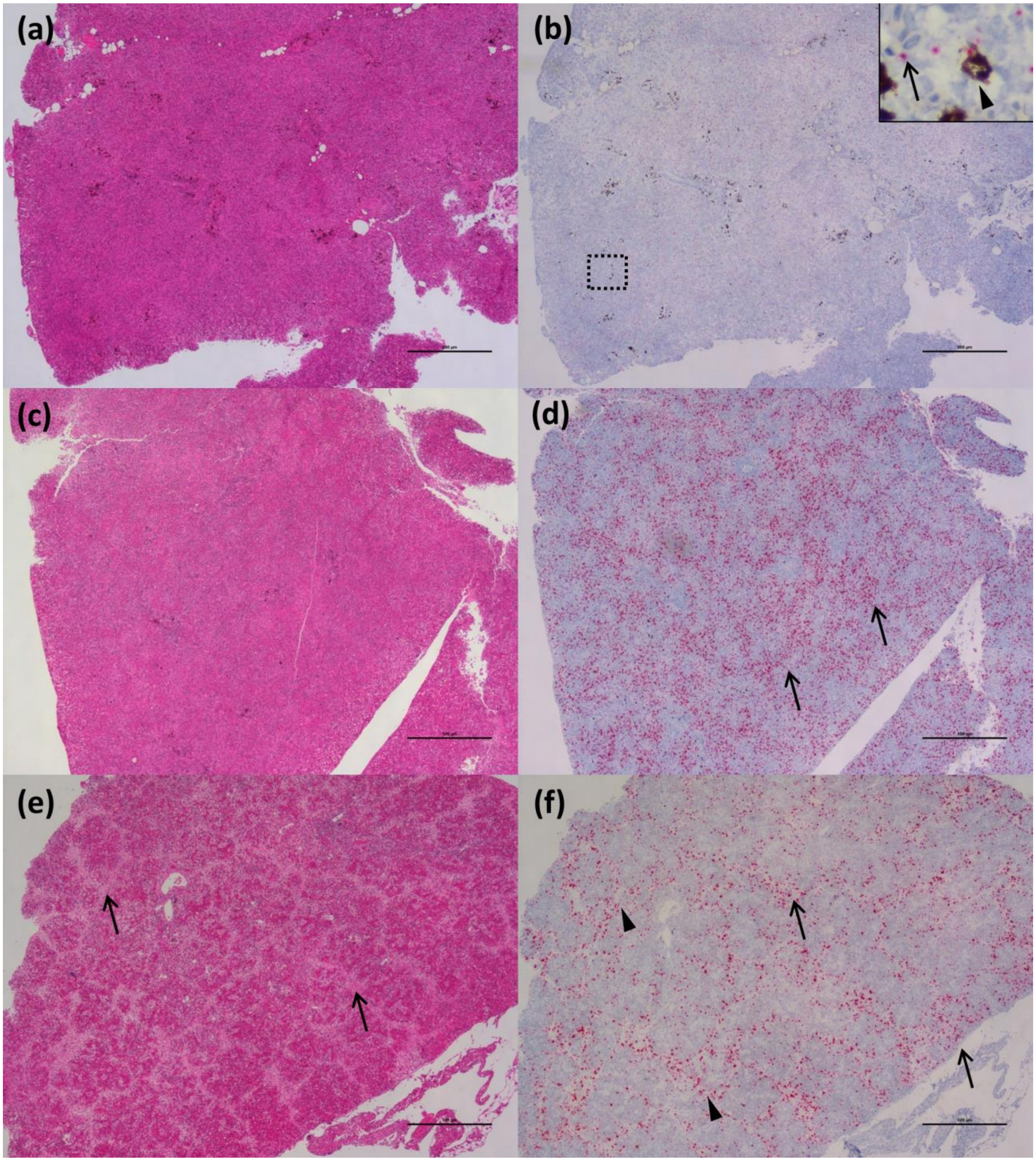

**Fig. 7.** Histopathology and ISH in Atlantic salmon spleen. Matching fields of consecutive sections from samples at various stages of HSMI disease development are stained with H&E (a, c, e) and ISH for PRV-1 L1 segment (b, d, f). The scale bar in all figures is 500µm. a) Group 2, with no apparent lesions. b) PRV (red) was localized in the RBCs (arrowheads), but also in some macrophages (arrowheads) (inset). c) Group 3, with no apparent lesions. d) A large number of PRV positive RBCs (red) were collected in the red pulp (arrows), along with PRV positive macrophages nearby the sinusoids. e) Group 4, showing mild congestion and moderate necrosis of the ellipsoids (arrows). f) A large number of PRV positive RBCs (red) were collected in the red pulp (arrows), although a larger number of damaged, PRV positive macrophages were observed in the ellipsoids (arrowheads).

## Chinook salmon

As the Chinook salmon used in our study were all from the DFO audit program, sampling only moribund and recently dead fish, we did not have true disease negative “controls” to use in our study. Hence, we restrict our analysis and description of histopathology and ISH to fish with molecular detection of PRV, which were further divided into groups depending upon whether they were classified as “VDD+” and whether they were diagnosed with jaundice/anemia (in which case they were all VDD+).

**Blood**: PRV was detected in the RBCs from all fish with molecular detection of PRV. The marking was characterized by several dots of various sizes inside the cytoplasm of the cells (Fig. 2b). At times, the dots were so numerous as to basically fill the entire cytoplasm of the RBCs. The marking in the RBCs was particularly evident in fish that were not classified as “VDD”.

**Heart**: No lesions in the heart were observed in fish that were not classified as “VDD+” (Fig. 8a), and in this fish PRV marking was restricted to the RBCs of the blood circulating inside the organ (Fig.8b). One fish (1/5) classifying as “VDD+” but not diagnosed with jaundice/anemia had mild endocarditis, localized exclusively in the spongy layer of the myocardium (Fig. 8c and 9a). In fish with “VDD+” alone, PRV was localized within some cardiomyocytes in the cytoplasm, mostly restricted to the spongy layer of the myocardium and often widespread in the tissue; a portion of these PRV-infected cardiomyocytes were associated with small nodules of inflammation (Fig. 8d and 9b). For fish diagnosed with jaundice/anemia, moderate or severe endocarditis in the spongy layer of the myocardium was identified in three out of nine fish, with sporadic inflammatory foci identified also in the compact layer. Overall, the inflammation was morphologically characterized by an infiltration of lymphocytes and macrophages, and was associated with focal areas of myocardial degeneration (Fig. 8e, 9c and 9e). In fish with jaundice/anemia, PRV was localized in the cardiomyocytes concentrated around regions with endocarditis in the spongy myocardium. Concurrently, a few cardiomyocytes of the compact layer also showed some dots of marking, and were usually associated with small aggregates of leucocytes (Fig. 8f, 9d and 9f).

The localization of both heart lesions and PRV marking reported in the samples affected by jaundice/anemia was remarkably different from that observed in Atlantic salmon affected by HSMI, as shown in Fig. S3.

**Liver**: One of three fish with molecular PRV detection but not classified as “VDD+” presented with severe hepatitis; this fish also had a confirmed case of rickettsiosis; the other two fish had no liver lesions (Fig. 10a). PRV was observed only in the RBCs in these fish (Fig. 10b). For fish classified only as “VDD+”, most (3/5) presented with widespread nodules and/or areas of moderate to severe necrosis, alongside broad areas of vacuolar degeneration (Fig. 10c). The same type of lesions (but often even more extensive) were also observed in fish diagnosed with jaundice/anemia (5/9), at times associated with inflammatory infiltrate (1/9) (Fig.10e). Two fish (* in Table 3) had no necrotic areas, but virtually all hepatocytes had varying degrees of vacuolar degeneration, characterized by a foamy cytoplasm. In fish classified as "VDD+", areas including cells affected by vacuolar degeneration and necrosis contained strongly positive PRV marking (Fig. 10d), however in fish diagnosed with jaundice/anemia, extended necrotic areas were positive for PRV, but to a lesser degree (Fig. 10f). Overall, in Chinook salmon positive for PRV and showing molecular signs of viral disease (VDD+), there was no distribution pattern of necrosis (as opposed to what observed in Atlantic salmon), and the type of lesions was more likely to be attributable to hypoxia due to anemia.

**Kidney**: Most fish in our study had varying degrees of interstitial hyperplasia (3/3, 4/5 and 8/9, respectively); mostly mild and moderate in fish not diagnosed with jaundice/anemia (Fig. 11a), but occasionally severe in fish diagnosed with jaundice/anemia. In fish not classified as "VDD+", PRV was localized in few RBCs as well as in some macrophages and melanomacrophages (Fig. 11b). Fish that were classified as "VDD+" or "VDD+ with jaundice/anemia" were also characterized by the presence of mild to severe necrosis of the renal tubules (3/5 and 3/9, respectively) (Fig. 11c and 11e), where there was also a substantial increase in the number of PRV positive macrophages. PRV marking were also found in the cytoplasm of damaged epithelial cells of the tubules and in the necrotic material inside the degenerate tubules (Fig. 11d). The occurrence of necrotic tubules and associated macrophage proliferation was intensified in fish diagnosed with jaundice/anemia (Fig. 11f). Four of nine fish diagnosed with jaundice/anemia also presented with mild to severe nephritis.

**Spleen**: No lesions were observed in the spleen of fish with molecular detection of PRV that were not classified as "VDD+" (Fig. 12a), for which PRV was present in only a few RBCs and reticular macrophages of the ellipsoids and around the sinuses (Fig. 12b). However, most of the fish classified as "VDD+" or "VDD+ with jaundice/anemia" presented with mild to moderate spleen congestion (3/5 and 4/9, respectively), mild to severe white pulp hyperplasia (2/5 and 4/9, respectively) and, in some cases, mild to severe necrosis of the ellipsoids (3/5 and 4/9, respectively) (Fig. 12c and 12e). As well, one fish diagnosed with jaundice/anemia showed signs of moderate hemosiderosis. The number of marked RBCs was considerably higher in fish classified as "VDD+", compared to those with no "VDD+" classification; these PRV-infected cells localized primarily in the sinusoids, which became engorged, distended and easily identifiable. An increase in positively marked macrophages around the sinusoids was also observed (Fig. 12d). In fish diagnosed with jaundice/anemia, the amount of PRV-infected RBCs and macrophages further increased to eventually occupy most of the organ, making it quite difficult to recognize the localization of the engulfed sinusoids within the red pulp, and reducing the white pulp to few, scattered nodules (Fig. 12f).

**Intestine**: No lesions in the intestine were observed in any of the Chinook salmon in our study (Fig. S4a, S4c and S4e), although positively marked RBCs were observed in all fish (Fig. S4b). Several enterocytes were also marked in fish classified as "VDD+" (Fig. S4d), but decreased in those diagnosed with jaundice/anemia (Fig. S4f).

A biplot PCA based on lesion scores from Table 3 depicts strong differential clustering between Atlantic and Chinook salmon along PC1 and PC2, respectively (Fig. 13). HSMI developmental stages in Atlantic salmon are well differentiated along PC1 (36.11% variation), while jaundice/anemia in Chinook salmon is highly loaded on PC2 (23.74% variation), although with more overlap between stages of disease development. Heart epicarditis and endocarditis, as well as myositis (muscle_itis), were the most differentiating lesions along PC1. Kidney interstitial hyperplasia was most strongly associated with PC2 (and jaundice/anemia), with contributions from splenic white pulpitis (spleen_wpu itis) and nephritis (kidney_itis). Liver necrosis was intermediate between PC1 and PC2, with higher levels associated with stronger disease development; however, this was a general category and the pattern of necrosis was notably different between the two species, as mentioned above. Necrosis of the splenic ellipsoids (spleen_ell nec) was also intermediate, increasing with disease development for both Atlantic and Chinook salmon. Barnard exact tests identified these same lesions as most differentiating between Atlantic and Chinook salmon within each stage of disease development (Table S2).

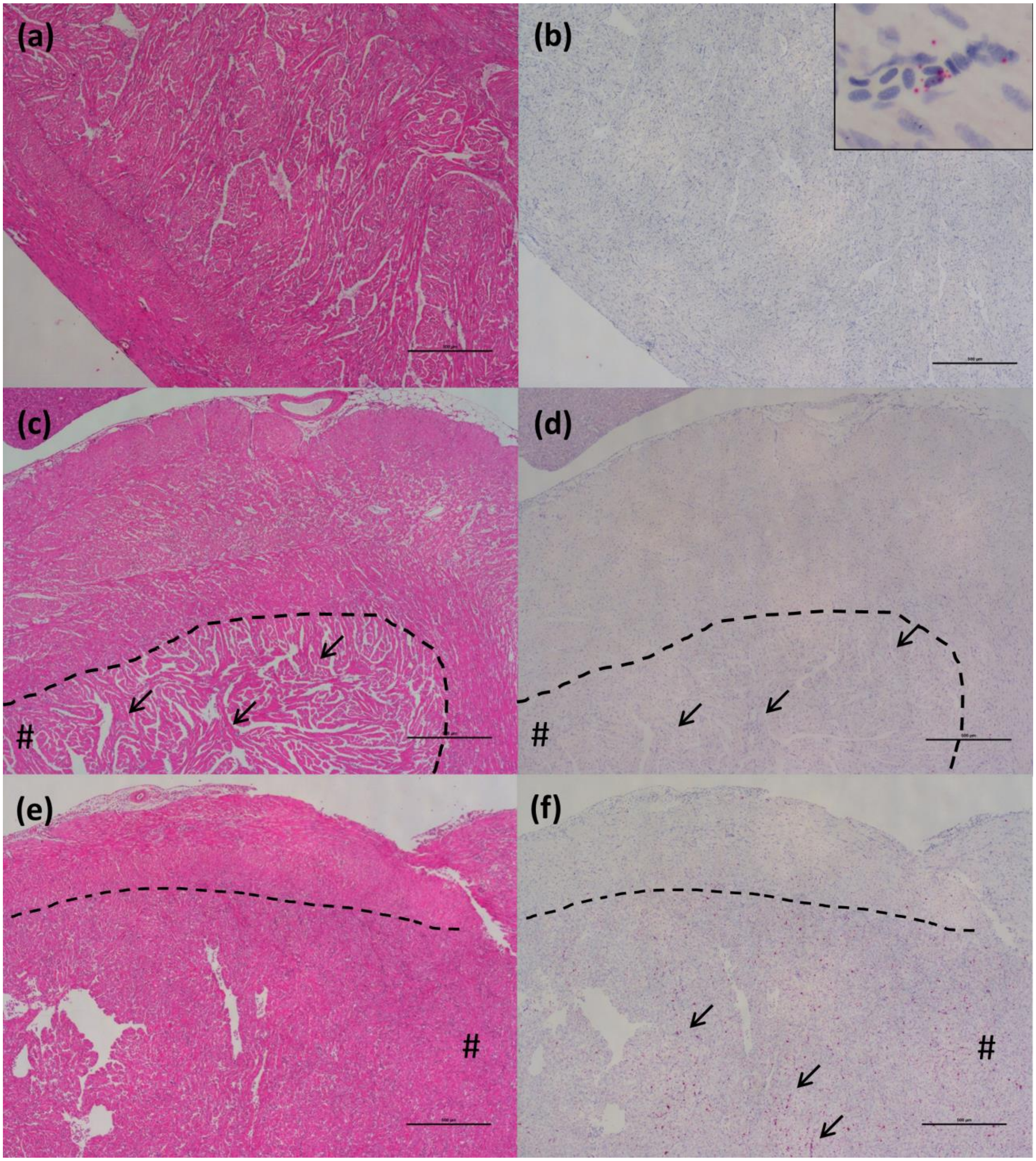

**Fig. 8.** Histopathology and ISH in Chinook salmon heart. Matching fields of consecutive sections from samples at various stages of jaundice/anemia development are stained with H&E (a, c, e) and ISH for PRV-1 L1 segment (b, d, f). The scale bar in all figures is 500µm. a) Group 1, with no apparent lesions. b) PRV (red) was localized in the RBCs only (inset). c) Group 2, showing small nodules of endocarditis (arrows) involving the spongy layer of the myocardium (hash). d) PRV positive cardiomyocytes (red) are localized in the spongy layer (arrows), in association with small inflammatory infiltrates. e) Group 3, showing severe endocarditis in the spongy layer of the myocardium (hash). f) Several PRV positive cardiomyocytes (red) were observed in the spongy layer of the myocardium (arrows), associated with severe endocarditis.

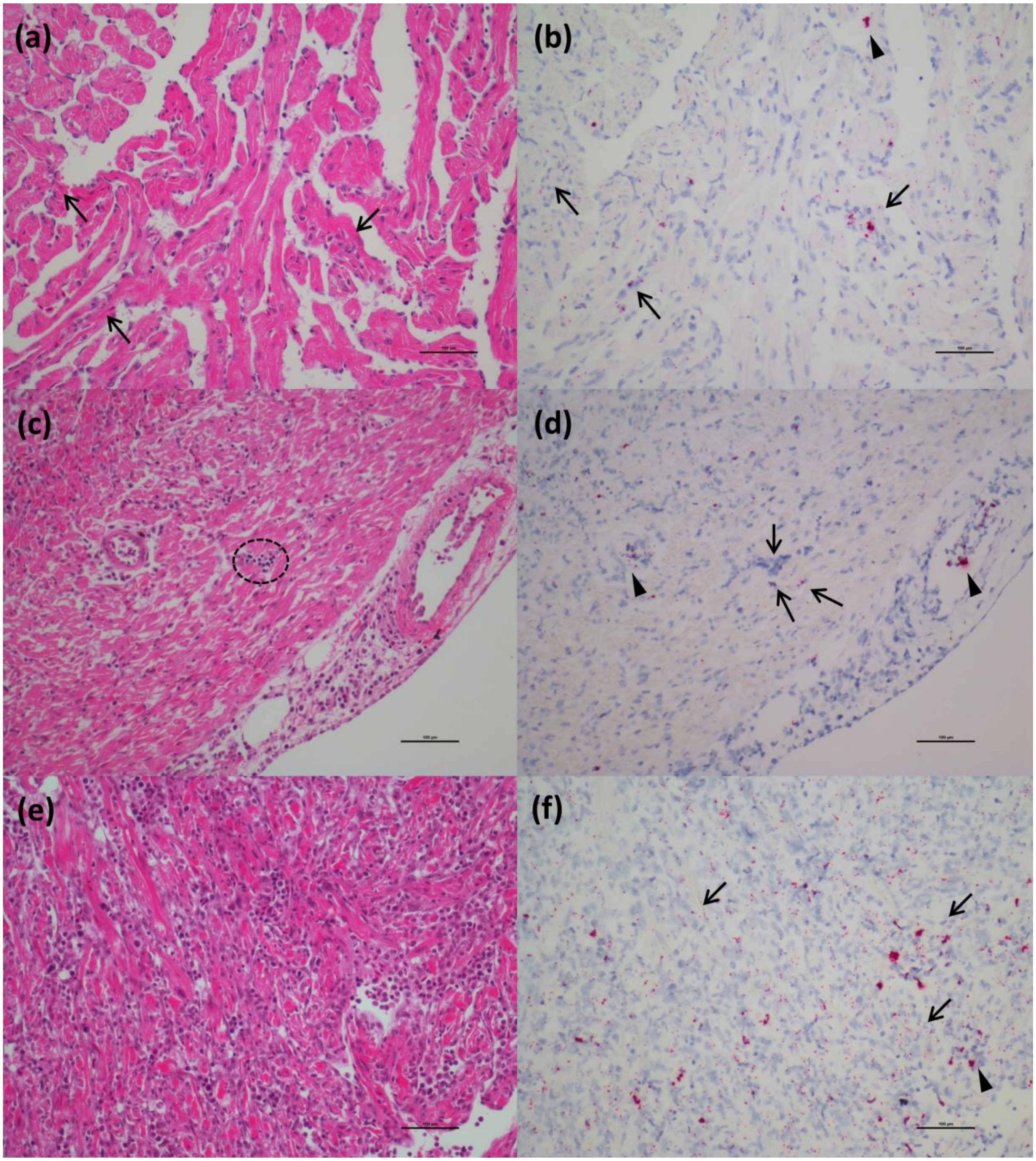

**Fig. 9.** Histopathology and ISH in Chinook salmon heart. Matching fields of consecutive sections from samples at various stages of jaundice/anemia development are stained with H&E (a, c, e) and ISH for PRV-1 L1 segment (b, d, f). The scale bar in all figures is 100µm. a) Group 2, showing small aggregates of leucocytes (arrows) surrounding cardiomyocytes in the spongy layer of the myocardium. b) PRV positive cardiomyocytes (arrows) are localized nearby the small nodules of endocarditis. PRV positive RBCs are also present in the lumen of cardiac cavity (arrowhead). c) Group 3, showing a small, single nodule of inflammation (circle) in the compact layer of the myocardium. d) PRV (red) was localized in cardiomyocytes nearby the inflammatory infiltrate (arrows). A few PRV positive RBCs (arrowheads) were present inside blood vessels. e) Group 3, showing diffuse, severe endocarditis in the spongy layer of the myocardium. f) PRV positive cardiomyocytes (red) are localized in the spongy layer (arrows). A few PRV positive RBCs are also present in the lumen of cardiac cavity (arrowhead).

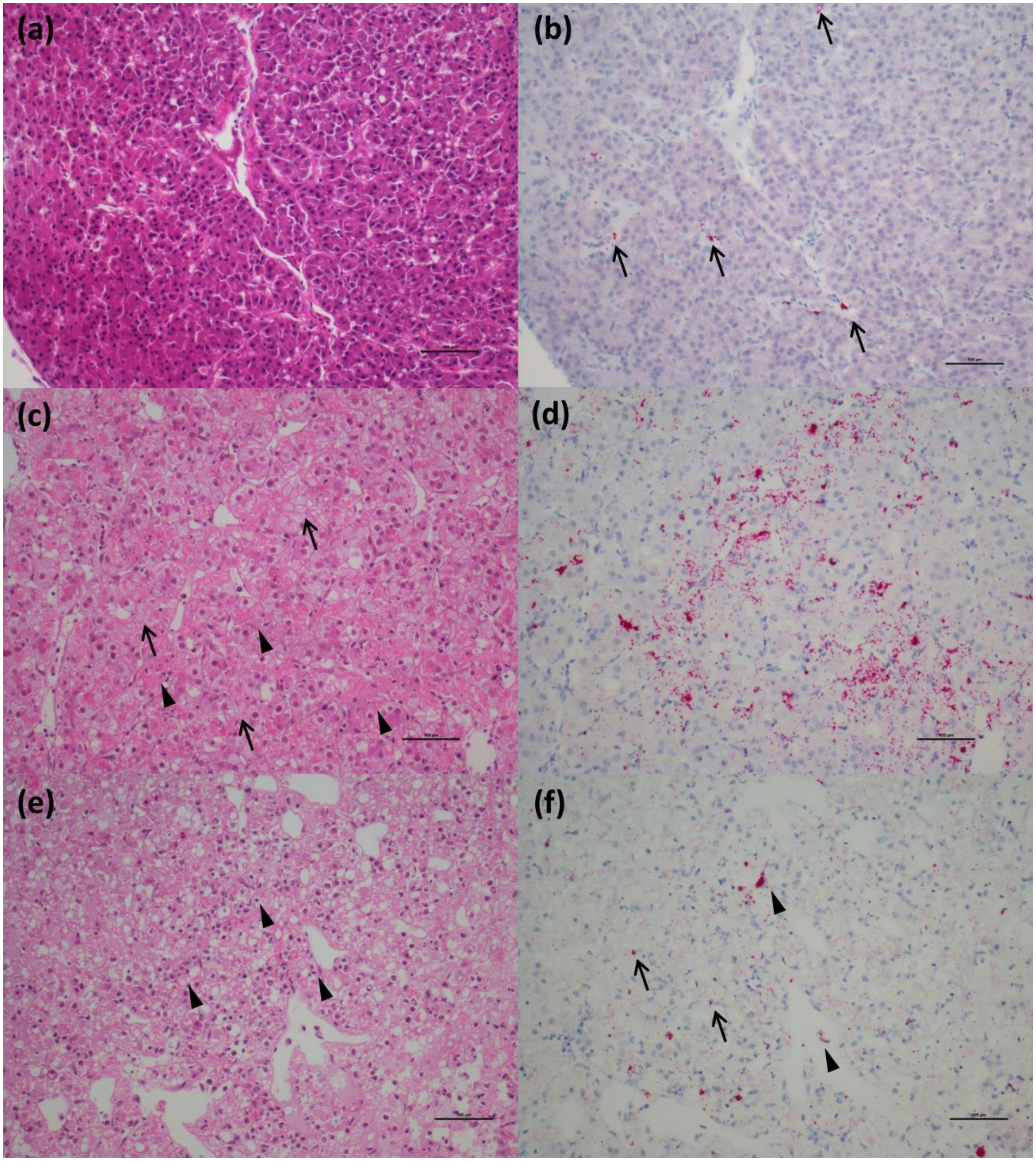

**Fig. 10.** Histopathology and ISH in Chinook salmon liver. Matching fields of consecutive sections from samples at various stages of jaundice/anemia development are stained with H&E (a, c, e) and ISH for PRV-1 L1 segment (b, d, f). The scale bar in all figures is 100µm. a) Group 1, with no apparent lesions. b) PRV (red) was localized in the RBCs (arrows) only. c) Group 2, showing areas of vacuolar degeneration (arrows) and necrosis (arrowheads). d) PRV (red) was diffusely present in degenerated and necrotic hepatocytes. e) Group 3, showing diffuse hepatic necrosis and few inflammatory cells (arrowheads). f) PRV (red) was localized mostly in the cytoplasm of the few, viable hepatocytes (arrows). A few RBCs were also positive to PRV (arrowheads).

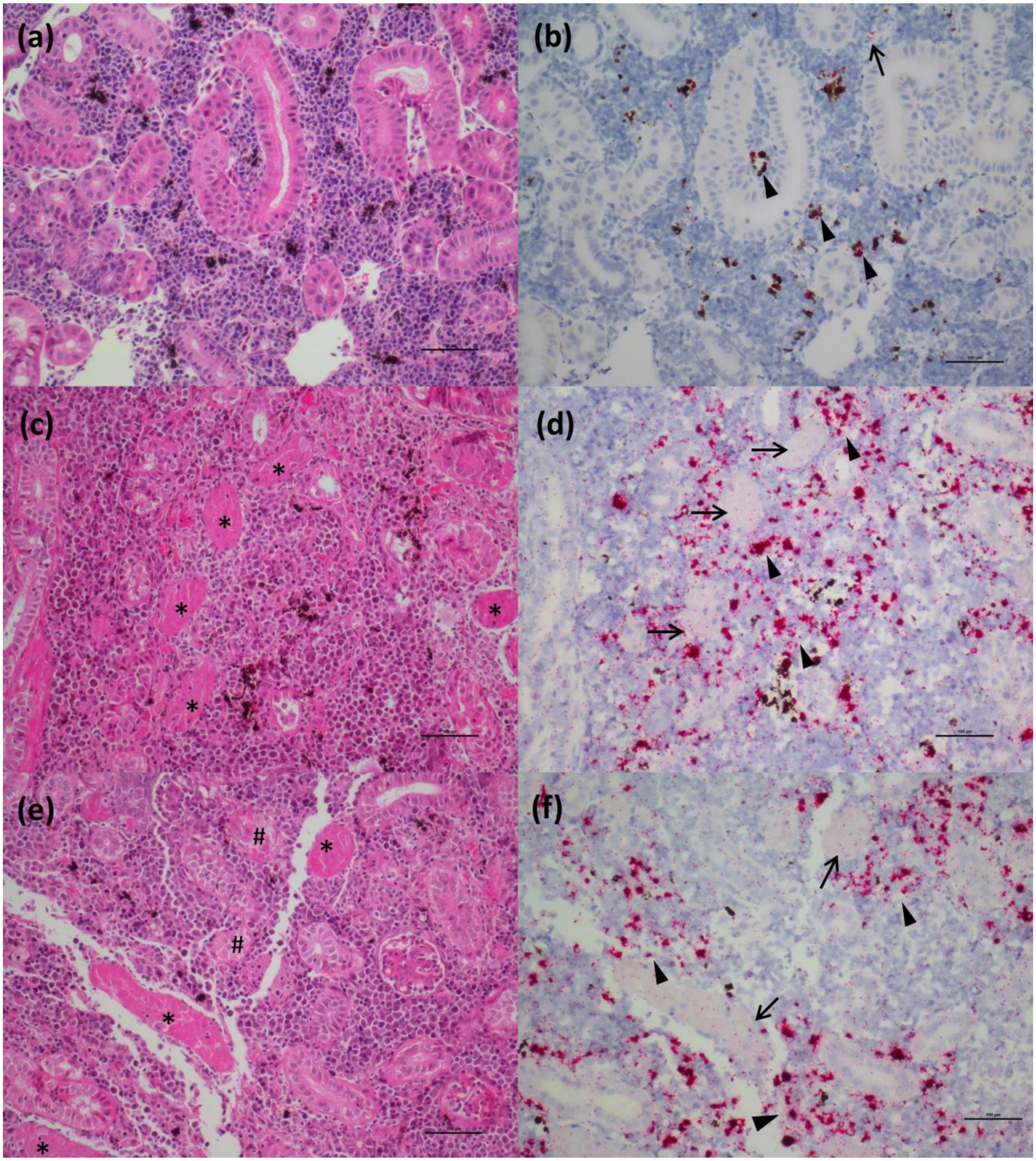

**Fig. 11.** Histopathology and ISH in Chinook salmon posterior kidney. Matching fields of consecutive sections from samples at various stages of jaundice/anemia development are stained with H&E (a, c, e) and ISH for PRV-1 L1 segment (b, d, f). The scale bar in all figures is 100µm. a) Group 1, showing mild interstitial hyperplasia. b) PRV (red) was localized in the RBCs (arrows), but also a few macrophages (arrowheads). c) Group 2, showing interstitial hyperplasia and several necrotic renal tubules (stars). d) PRV (red) was mostly localized in the peritubular (renal portal) macrophages (arrowheads), but viral particles were also present in the material contained inside the necrotic tubules (arrows). e) Group 3, showing interstitial hyperplasia and several necrotic (stars) and damaged (hash) renal tubules. f) PRV (red) was localized mostly in the macrophages (arrowheads), as well as in the material contained inside the necrotic tubules (arrows).

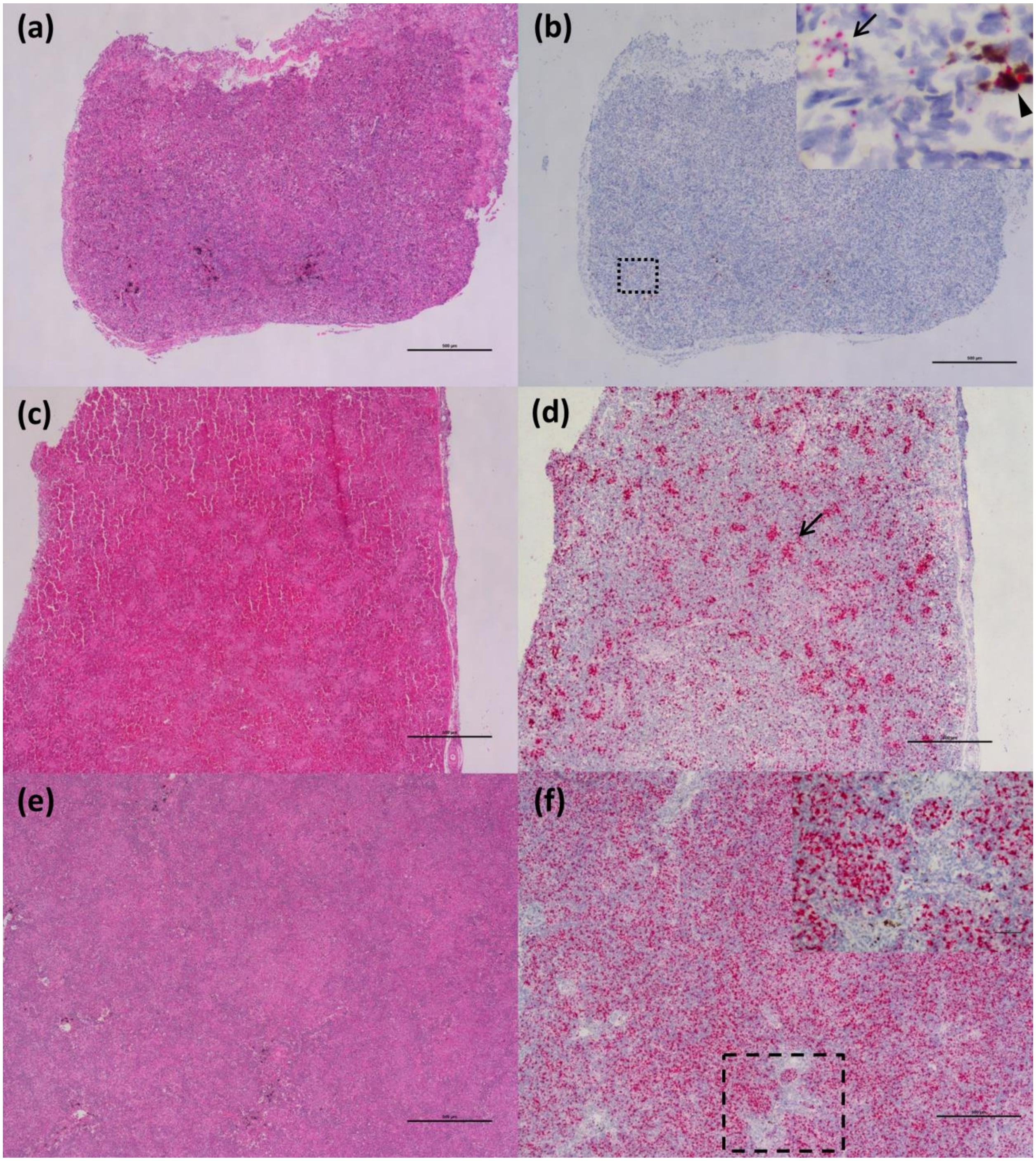

**Fig. 12.** Histopathology and ISH in Chinook salmon spleen. Matching fields of consecutive sections from samples at various stages of jaundice/anemia development are stained with H&E (a, c, e) and ISH for PRV-1 L1 segment (b, d, f). The scale bar in all figures is 500µm. a) Group 1, with no apparent lesions. b) PRV (red) was localized in the RBCs (arrows), but also in some macrophages (arrowheads) (inset). c) Group 2, showing increased congestion and white pulp hyperplasia. d) A large number of PRV positive RBCs (red) were aggregated in the red pulp, highlighting the position of the sinusoids (arrows), along with PRV positive macrophages nearby the sinusoids. e) Group 3, showing blood congestion. f) PRV (red) was localized in virtually any RBCs and macrophages of the red pulp.

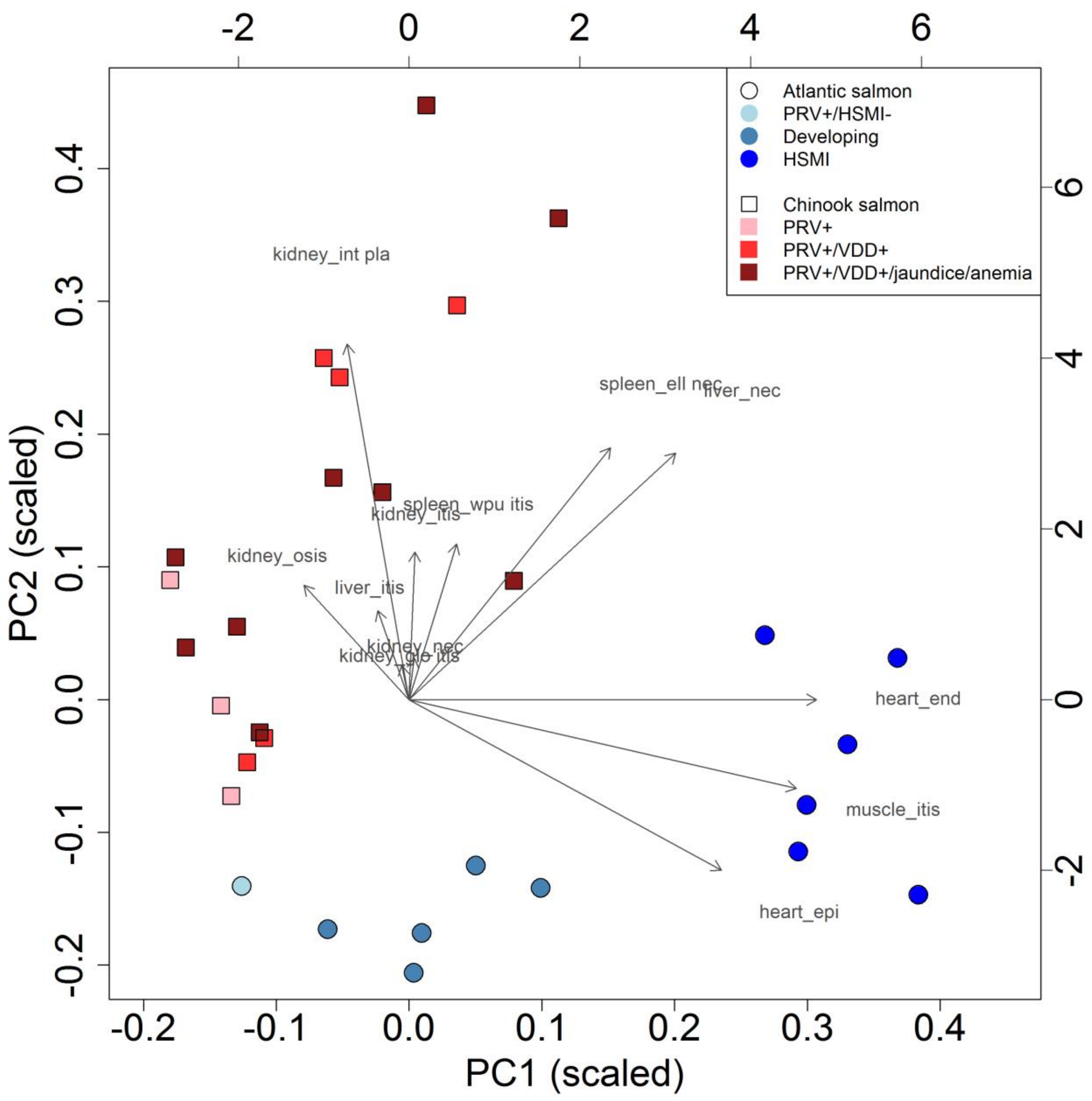

**Fig. 13.** Biplot for PCA of 35 samples (18 Atlantic and 17 Chinook) and 12 lesions (8 inflammatory and 4 necrotic; necrotic lesions: 'liver_nec', 'spleen_ell nec', 'kidney_osis' and 'kidney_nec"). The number displayed next to two points representing Atlantic salmon indicates the number of fish that have the exact same lesion scores (and therefore have the same PC scores and are plotted on top of each other). The bottom and left axes show sample scores for PC1 and PC2 scaled to unit sums-of-squares (colored points and squares) while the top and right axes show variable loadings scaled by the standard deviation of the principal components times the square root of the number of samples.

## Immunohistochemistry with antibody to hemoglobin

The immunohistochemistry carried out against hemoglobin subunit beta demonstrated positive marking in the RBCs of all samples, as expected. However, all samples belongings to the “PRV+/VDD+” group and the “jaundice/anemia” group (6 out of 6) tested for hemoglobin also showed positive marking in the material (renal casts) contained in the lumen of the necrotic renal tubules (where present), and some hemoglobin-positive material was found in a few apparently healthy epithelial cells of the renal tubules (Fig. 14a-d). This finding was totally absent in all fish belonging to the “PRV+ only” group (2 out of 2) and in the negative control samples.

Positively marked hepatocytes were also observed in the samples belonging to the “PRV+/VDD+” and “jaundice/anemia” groups. In the “PRV+/VDD+” group, some hepatocyte presented with intracytoplasmic hemoglobin-positive vacuoles (Fig. 14e). In “jaundice/anemia” group, almost any hepatocyte presented with some degree of marking to hemoglobin, in particular cells identified by apoptotic nuclei and associated with a large, clear intracytoplasmic vacuole (Fig. 14f).

## PRV Phylogenetic Sequence analysis

Phylogenetic analysis of the S1 segment including both the PRV-2 and PRV-3 strains shows that the divergence between sequence variants within the PRV-1 strain is low compared to the evolutionary distance between PRV-1, 2 and 3 (Fig. 15). Based on the S1 segment, all BC sequence variants fell within a single phylogenetic cluster within PRV-1, which also included sequences from Norway (Atlantic salmon), and Chile (Coho salmon). Although PRV-1 has previously been broken down into sub-strains PRV-1a and 1b, this distinction is not helpful considering that there are sequences which lie somewhere between the two sub-strains (e.g. JN991012.1, JN991007.1), and further sequencing of the PRV-1 quasispecies will likely see these sub-strains break down. Within PRV-1, the single Norwegian reference sequence utilized was the most divergent, with BC strains generally grouping together (Fig. S5). From the available data, there is more variation within PRV-1 in Norway than in BC, where, despite infection in different species, there is little variation between samples.

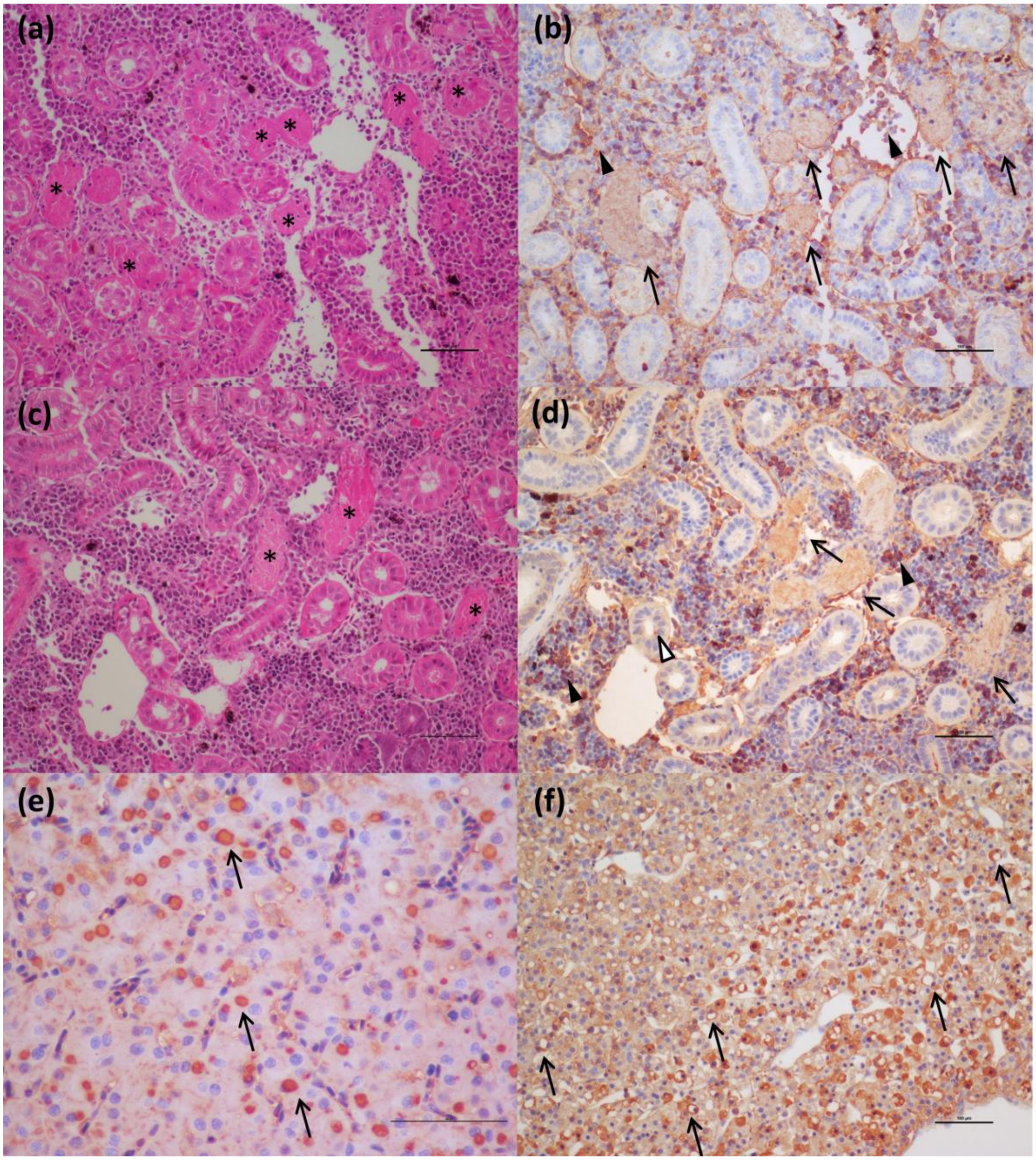

**Fig. 14.** Histopathology and immunohistochemistry in Chinook salmon posterior kidney and liver. Matching fields of consecutive sections from kidney samples at various stages of jaundice/anemia development are stained with H&E (a, c) and IHC for hemoglobin (b, d). The scale bar in all figures is 100µm. Two liver samples stained with IHC for hemoglobin (e and f) from two different stages of jaundice/anemia development. Scale bar in all samples is 100µm. a) Kidney of fish belonging to group 2, showing necrotic material (renal casts) in lumen of necrotic renal tubules (stars). b) Hemoglobin (brownish red) is observed in RBCs (arrowheads) and in the material contained in the necrotic renal tubules (arrows). c) Kidney of fish belonging to group 3, showing necrotic material (renal casts) in lumen of necrotic renal tubules (stars). d) Hemoglobin (brownish red) is observed in RBCs (arrowheads), in the material contained in the necrotic renal tubules (arrows) and as inclusions in the epithelial cell of some apparently healthy renal tubules (open arrowhead). e) Liver of fish belonging to group 2, showing widespread, intracytoplasmic hemoglobin-positive vacuoles (brownish red – arrows). f) Liver of fish belonging to group 3, showing diffuse hemoglobin marking (brownish red), and particularly concentrated in hepatocytes presenting with apoptotic nucleus and a large, clear, intracytoplasmatic vacuole (arrows).

The Bayesian inference of phylogeny of PRV sequences from Atlantic and Chinook salmon from this study did not show consistent differences among sequence variants for each segment (Fig. S5). For each sample, the sequence variation within segments was remarkably low when compared with the Norwegian sequence. Nucleotide sequence identities among PRV-1 sequences in BC were >99%, with minor variance among segments (minimum identity of 99.4; S1). Alternatively, when PRV-1 was compared with Strains PRV-2 and PRV-3, the nucleotide identities were 73.4% and 80.1%, respectively (Dhamotharan et al. 2018), whereas the minimum sequence identity within PRV-1 was >95.1% in the S1 segment (Godoy et al. 2016).

Although some trees depict a degree of differential clustering among salmon species (e.g. M1, M2), close inspection of the multiple sequence alignment (not shown) did not reveal any SNPs consistently different between species if the reference genome from Norway was included in the analysis (see Fig. S5 for all trees). Perhaps the most distant clustering between sequences derived from Atlantic and Chinook salmon occurred for the M1 segment, which was supported by a posterior probability of 0.999.

For Chinook salmon, we sequenced PRV from the west and east coasts of Vancouver Island, and found that for all but segment S1, the west coast samples (labeled P2) clustered tightly together (Fig. S5). However, all four west-coast samples were from the same company, with farms in close proximity to one another. In contrast, the Atlantic salmon were from farms on the east coast of Vancouver Island (A3.2 and A3.3) and the central coast of BC (A3.5), there was no obvious geographic pattern in sequence clustering, nor a pattern associated with individual companies (data not shown). We also did not observe particularly close clustering of sequences from fish sampled from the same farm three months apart (samples G460 and G383), nor from those taken from different year-classes within the same farm (G808 and G609). However, PRV-1 from a semi-close-contained Chinook salmon farm (G444) was generally the most divergent from other Pacific salmon sequences.

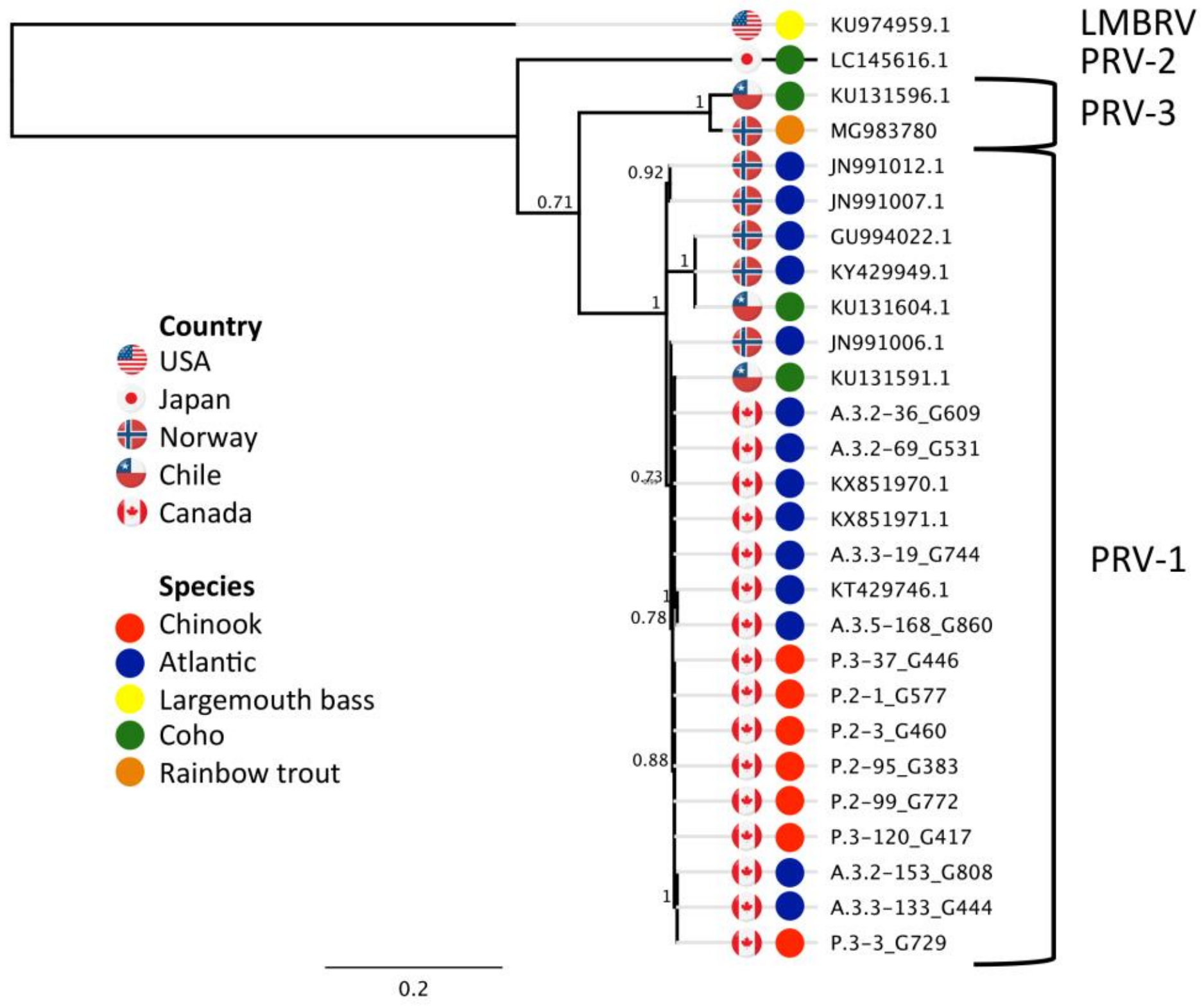

**Fig. 15.** Bayesian inference of phylogeny of the PRV S1 segment comparing sequence variants from Atlantic and Chinook salmon from farms containing fish diagnosed with HSMI or jaundice/anemia, respectively as well as representatives of all PRV strains identified worldwide. GU994017.1 is the first PRV sequence identified in Palacios et al. (2010). KY429949.1 is the PRV-1 sequence for the virus demonstrating a cause and effect relationship with HSMI (Wessel et al. 2017). The Bayesian support values (posterior probability) are labelled and the scale bar represents number of nucleotide substitutions per site. Largemouth bass reovirus (LMRV) was included as an outgroup. PRV genome sequences were deposited to GenBank under accession numbers MH093900 - MH094029.

# Discussion

This is the first study to specifically localize PRV across multiple tissues during disease development in salmon. We demonstrate an intimate relationship between viral localization and development of histological lesions that supports an etiological role for the virus in the development of both HSMI in Atlantic salmon and jaundice/anemia in Chinook salmon. It is already established that Strain PRV-1 causes HSMI in Atlantic salmon (Wessel et al. 2017), and that Strain PRV-2 causes EIBS in Coho, a jaundice/anemia-related disease in Japanese Coho salmon (Takano et al. 2016). Moreover, strains of PRV have been associated with other diseases. For example, PRV-1 has been associated with jaundice syndrome in farmed Chinook (Miller et al. 2017) and Strain PRV-3 with HSMI-like disease and anemia in Rainbow trout in Norway (Olsen et al. 2015) and Coho salmon in Chile (Godoy et al. 2016). Therefore, all the three recognized strains of PRV have now been affiliated with a jaundice-related disease in three species of Pacific salmon, and weight of evidence worldwide, and our findings in BC, suggests that PRV-1, the only PRV strain detected in BC salmon, likely causes both diseases, HSMI and jaundice/anemia, in Atlantic and Pacific salmon respectively.

Full viral genome sequencing of PRV in Chinook salmon diagnosed with jaundice/anemia or classified as PRV+/VDD+, sampled from seven farm audits collected over three years, and in Atlantic salmon diagnosed with HSMI, sampled over three years from six farm audits distributed across the BC industry, validated that only closely related sequence variants within PRV-1 occur on BC farms. The sequence variants of PRV-1 from BC share high sequence identity to sequences in Norway (maximum 99.7% identity [JN991006.1]) and Chile (maximum 99.7% identity [KU131591.1 from Coho]) for S1, which is within the range of variation observed in BC. Over the entire viral genome maximum sequence identity between BC variants of PRV-1 and the first PRV-1 sequence described by Palacios et al. (2010) was not as high, and varied among segments from 96.6 (S1) to 99.2 (M1). S1 and M2 are the two segments showing strongest divergence between this Norwegian sequence and BC PRV-1 variants (96.6 and 96.8 percent respectively). In *Mammalian orthoreovirus* (MRV) these segments code for proteins that confer virulence in mice (Weiner et al. 1977; Rubin and Fields 1980), with S1 coding for the cell-attachment protein, which is important for neurovirulence and tissue tropism (Weiner et al 1977). If these proteins have similar functions in *Piscine orthoreovirus*, it is possible that the minor differences observed between the Palacios Norwegian and BC sequence variants of PRV-1 could result in shifts in virulence. This is an area that will be examined in future challenge studies.

While there were no consistent differences in PRV-1 sequences between Atlantic salmon with HSMI and Chinook salmon with jaundice/anemia, sequences from Atlantic salmon clustered separately from Chinook salmon for the L1, M1, and M2 segments; however, there were no single-nucleotide or amino-acid substitutions that were consistently differential between salmon species (data not shown). Hence, the low sequence divergence within PRV-1 is consistent with the virus cross-infecting Atlantic and Chinook salmon, and suggests that the virus can transmit diseases between salmon species.

The molecular viral disease development panel, or VDD, was developed in the Miller et al. (2017) study from meta-analysis of multi-cohort microarray data from viral disease challenge studies for five diseases caused by RNA viruses (infectious hematopoietic necrosis [IHN], infectious pancreatic necrosis [IPN], infectious salmon anemia [ISA], cardiomyopathy syndrome [CMS] and heart and skeletal muscle inflammation [HSMI]), and validated using RNA samples from independent challenge (IHNV across multiple species) and field (HSMI in Atlantic salmon and jaundice/anemia in Chinook salmon) samples. In addition to differentiating carrier and disease states for viruses, the VDD panel could also differentiate bacterial and viral disease states. It was shown that a robust set of eleven VDD biomarkers was all that was required; these included SRK2, DEXH, RSAD, UBL1, IFIT5, NFX, HERC6, VHSVIP4, IF44A, CD9, and MX. The application of the VDD panel was particularly useful in understanding the disease development process, and identifying fish at an early stage of disease progression, before clinical (and sometimes morphological) evidence of disease is recognized. In applying this tool across both Atlantic and Pacific salmon, we demonstrated that almost all fish diagnosed specifically with PRV-associated disease (HSMI or jaundice/anemia) showed a strong host VDD response. Exceptions were two Atlantic salmon with low PRV load that we surmise may have been in a recovery stage whereby lesions persist after the virus is all but cleared, as demonstrated in Norwegian studies (Finstad et al. 2012). Fish specifically diagnosed with disease account for only 17% of Atlantic salmon and 41% of Chinook salmon with PRV loads > $10^4$ copies per µg nucleic acids classifying as VDD+, suggesting that many more fish may have been in a pre-clinical disease state. However, in both species, there are also fish (33% for Atlantic, 19% for Chinook salmon) that contain PRV loads > $10^4$ copies per µg nucleic acids that do not classify as VDD+. Our research also shows that in these fish, the virus was not observed outside of the RBCs, the primary infective cells for PRV. These data support the hypothesis put forward by Finstad et al. (2014) and Wessel et al. (2015) that PRV is highly tolerated by the salmon host (Atlantic salmon) when replicating exclusively in RBCs, and that the host elicits no strong immune reaction or pathological response as long as the virus remains in RBCs (classified in our study as PRV+ only). This explains what has been considered an anomalous finding that fish with high loads of the virus may or may not show any indication of disease, which led some scientists to speculate that PRV is not a primary cause of disease (Garver et al. 2016), later refuted in the study by Wessel et al. (2017) demonstrating a cause and effect relationship between PRV-1 and HSMI.

Through application of the molecular VDD tool, and differentiation of fish in a PRV+ carrier state from those at early stages of disease development, we also learn that this anomaly is not as strong in Chinook salmon as it is in Atlantic salmon. All Chinook salmon in our study with mixed tissue PRV $>10^5$ copies per µg nucleic acids classified as VDD+ and showed evidence of RBC rupture, leading to release of virus and high levels of hemoglobin into the tissues. These data are consistent with an upper limit of load tolerance to infection of RBCs in Chinook salmon that is not observed in Atlantic salmon which that may result in a higher susceptibility of Chinook salmon to disease development. Ideally, an exact tolerance threshold value needs to be determined experimentally, specifically monitoring loads in RBCs, but the use of regulatory audit fish, that were moribund or dead upon sampling, precluded this possibility.

Our ISH data indicate that RBC lysis and release of virus in Chinook salmon results in the elicitation of an immediate strong 'viral disease' host response (VDD). Additionally, we observed hemosiderin appearance in the spleen of some fish (evidence of metabolized hemoglobin from ruptured RBCs), and massive erythrophagocytosis of damaged/ruptured RBCs (containing PRV) and possibly free virus in spleen and kidney. Immunohistochemistry analysis for hemoglobin also confirmed the presence of high level of hemoglobin in the liver (particularly in damaged/apoptotic hepatocytes) and kidney (in the renal casts resulting from renal tubules necrosis) of fish diagnosed with jaundice/anemia or even just PRV+/VDD+. High levels of hemoglobin, bilirubin and heme byproduct are known to be toxic for the liver cells (Sakai et al. 1994) as well as for the epithelial cells of the kidney tubules (Tracz et al. 2007; Deuel et al. 2016). These are similar disease characteristics to comparable PRV-related diseases in Pacific salmon in Chile, Japan and Norway (Sakai et al. 1994; Smith et al. 2006; Olsen et al. 2015; Godoy et al. 2016; Takano et al. 2016). Moreover, mild to severe myocarditis primarily involving the spongy layer of the myocardium (as opposed to the whole heart, including mostly the epicardium and compact layer of the myocardium, in HSMI) occurs transiently in Chinook salmon. This pattern of inflammation has also been observed in PRV-associated diseases in Coho salmon and Rainbow trout (Hayakawa et al. 1989; Olsen et al. 2015; Godoy et al. 2016).

Interestingly, many of the Chinook salmon classified as PRV+/VDD+, but not with jaundice/anemia, pathologically showed advanced states of disease, suggesting that clinical signs (i.e. jaundice and anemia), which were principally used to classify fish in our study, may occur quite late in the disease process, potentially resulting in low diagnostic sensitivity on farms. Moreover, at the peak of the clinical manifestation of the disease, the virus is not always prevalent in the heart, and myocarditis lesions are not always present, suggesting a temporary occurrence of these lesions. By contrast, extensive areas of hepatonecrosis and kidney tubule necrosis are present in regions with high abundance of virus. Whether these degenerative/necrotic lesions are the result of viral infection specifically in the cells that become necrotic, or through the toxic levels of heme being collected, processed and accumulated by these tissues, is not clear, but in either case, these changes are clearly associated with PRV-1 virus-associated hemolysis.

ISH analysis of the HSMI disease development process in Atlantic salmon, known to be caused by PRV-1 infection, showed a strong association of the virus within regions and cells showing tissue damage. In the “developing” stage, the virus begins to move outside of RBCs and infect cardiomyocytes, mostly in the compact layer of the myocardium, but also, to a lesser extent, in the spongy layer, as was also indicated through immunohistochemistry by Wessel et al. (2017). In this early stage of disease, several cardiomyocytes become infected. The cardiomyocytes damaged by the intracellular activity of the virus induce the inflammatory reaction of the host, identified as foci of inflammation characterized by the presence of lymphocytes and macrophages. When the fish were diagnosed with HSMI, high quantity of the virus, and associated inflammatory lesions, spread throughout the heart, resulting in panmyocarditis by involving both compact and spongy layers of the myocardium, and sometimes even the atrium.

While our study did not specifically investigate ISH in skeletal muscle lesions (as we had difficulty breaking up muscle tissue sufficiently to allow our RNA probe to enter without causing the tissue to become over-digested or detach from the slide), we have described the temporal patterns of inflammatory lesions in skeletal muscle that track those in the heart (Di Cicco et al. 2017), and include these data herein. However, in Atlantic salmon, we investigated for the first time the viral localization in tissues other than the heart. Viral distribution in the spleen was similar to that in Chinook salmon in the early stages of disease, also showing erythrophagocytosis, but it didn’t reach the massive engulfment of damaged RBCs during the peak of the disease as demonstrated in Chinook salmon. Moreover, unlike in Chinook salmon, there was no evidence of necrotic lesions in liver (other than those described as “heart failure” pattern of necrosis) or kidney in most fish despite a moderate prevalence of the virus in hepatocytes and renal macrophages

One common finding between HSMI and jaundice/anemia was the localization of PRV particles in the intestine (i.e. enterocytes), particularly during the early stages of disease development, when a high amount of virus is free in the blood. This finding suggests and confirms that the intestine can act not only as an entry point for the virus, but also as a possible route for virus dissemination in the environment, as previously indicated by Hauge et al. (2016).

Our study was able to illustrate that the key difference in the PRV-associated disease manifestation in Atlantic and Pacific salmon probably lies in the difference between the ability of the species to tolerate high loads of the virus in RBCs. In HSMI, there is no evidence of RBC lysis (also, no anemia clinically reported), but rather PRV appears to leak from RBCs upon reaching a high viral load. We hypothesize that at that stage, the virus moves into the plasma and infects other cell types, with particular predilection for cardiomyocytes and (to a lesser extent) myocytes of the skeletal muscle. Not being able to support viral replication, these cells undergo structural changes that expose the virus itself or antigenic cellular components that eventually trigger the inflammatory reaction within these tissues. This type of reaction is cell mediated (CD8+ and macrophages) as demonstrated by Mikalsen et al. (2012), characterizing a typical Type IV hypersensitivity reaction. If inflammation becomes severe, it can result in heart failure and death.

By contrast, there is consistent evidence of RBC rupture (e.g. anemia, hemosiderin, enhanced macrophages-erythrophagocytosis, hemorrhaging, excess of free hemoglobin, excess of free bilirubin and jaundice) in PRV-associated diseases in Pacific salmon around the world (Sakai et al. 1994; Olsen et al. 2015; Godoy et al 2016; Takano et al. 2016), suggesting that PRV is released "*en masse"* during RBC lysis. This would also result in large quantities of hemoglobin being released into the tissues. Hemoglobin is catabolized and removed by hepatocytes, which can be overwhelmed by toxic levels of the breakdown product heme, and the yellow compound, bilirubin, byproduct of heme catabolism. Exposure to toxic levels of these byproducts can result in necrosis of the hepatocytes (Sakai et al. 1994), degeneration and necrosis of the epithelial cells of kidney tubules (Tracz et al. 2007; Deuel et al. 2016), leading to a yellow appearance of the internal organs (especially the liver) and in extreme cases the epithelial cells on the body of the fish. These lesions can also potentially lead to liver and kidney failure, and eventually death. Anemia, reported in all Pacific salmon species affected by strains of PRV (Olsen et al. 2015; Godoy et al 2016; Takano et al. 2016), also results from a massive reduction in RBCs that could lead to a reduced oxygen uptake capacity and hence a substantially higher short-term morality rate. Hence, the occurrence of transitory lesions in the heart seems to be characterized by a pathogenic mechanism different to the one affecting liver and kidney. The type of lesions reported for such organs also support this hypothesis, being inflammatory processes primarily involved in the heart (just like HSMI, despite a different viral localization/tropism in Chinook), and degenerative/necrotic processes in liver and kidney.

Given that *Piscine orthoreovirus* Strain PRV-1 causes HSMI in farmed Atlantic salmon, and likely causes jaundice/anemia in farmed Chinook salmon, the absence of consistent sequence variation within PRV-1 in diseased Atlantic and Chinook salmon implies a risk of transmission of the virus from farmed salmon to wild Pacific salmon. As Atlantic salmon accounts for 97% of the total biomass of farmed salmon in BC, answering the question of whether the virus causing HSMI in Atlantic salmon causes disease in Pacific salmon is critical when assessing risk. PRV-1 occurred in 65-75% of farm audit samples of Atlantic and Chinook salmon, with approximately 25% of the salmon having high viral loads. In farms where HSMI disease ensues, BC Atlantic salmon show a morbidity rate (i.e. presence of histological lesions) of >80%, similarly to that reported in Norway (Kongtorp et al. 2004a; Kongtorp et al. 2006), and retain high loads of the virus and evidence of inflammatory lesions for a prolonged period of time (in the study reported by Di Cicco et al. (2017), over 11 months). Hence, the environmental footprint and potential for viral transmission to wild salmon from disease-impacted farms is likely high. While neither disease has yet been identified in wild fish to date, given the high sensitivity of the molecular viral disease development tool to early developing disease states, and of ISH to localize the virus across tissues, we can now apply these approaches to empirically determine whether any evidence exists that PRV-associated diseases may be occurring in migratory salmon in BC.

HSMI is one of the most common infectious diseases of Norwegian farmed salmon (Garseth et al. 2017). Moreover, disease modeling has identified that fish farming intensity in a region is a major risk factor for HSMI outbreaks (Kristoffersen et al. 2013; Morton et al. 2017), suggesting that water-borne transmission may be an important factor in the spread of disease (Hjeltnes et al. 2016). These data also suggest that farming intensity could enhance exposure of migrating smolts to PRV (Garseth et al. 2017). Escaped farmed salmon, which are most often infected with PRV, could also be a transmission vector for freshwater infections in wild fish if they enter rivers (Madhun et al. 2015; Hjeltnes et al. 2016, 2017). Importantly, there have been recent reports of HSMI outbreaks occurring in freshwater hatcheries in Norway, which represents a shift in what was considered until recently a disease restricted to sea net pens (Hjeltnes et al. 2016, 2017). There is some indication as well, that PRV infection in freshwater may be associated with earlier, and more significant HSMI outbreaks in sea net pens, and as a result, some farmers are attempting to produce PRV-free smolts (E. Rimstad, personal communication, 2017; Hjeltnes et al. 2017).

Prevalence of PRV is low (<3%) in wild migrating BC Chinook and Sockeye salmon smolts that have not been exposed to salmon farms (Siah et al. 2015; Morton et al. 2017; Purcell et al. 2018; Tucker et al. 2018; Miller, unpublished data). Higher prevalence is observed in fall, especially in Chinook salmon on the west coast of Vancouver Island that remain resident in the same bays and inlets occupied by salmon farms for prolonged periods of time (up to a year) (Miller, unpublished data). An epidemiological study is currently underway to assess the role of salmon farms in the transmission of PRV to wild stocks, based both on patterns of prevalence distributions and full genome sequencing of the virus across farmed and wild populations over multiple years and geographic locations.

## Acknowledgements

This research was supported by funding for the “Strategic Salmon Health Initiative” provided by the Pacific Salmon Foundation (PSF), Genome British Columbia, and Fisheries and Oceans, Canada, co-led by KMM and Dr. B. Riddell of the PSF. We thank Dr. Riddell for his tireless support of this program and funding of ED and multiple individuals within the DFO audit program, including Dr. M. Shepherd, Dr. I. Keith, H. Manchester, T. Heiman, F.K. Khavar, and M. Charbonneau, for facilitating the collection and transfer of samples to our program. We also thank Marine Harvest and Cermaq for agreeing to provide and assist in the collection of samples from their farms, as well as the veterinarians Dr. Larry Hammell and Dr. Megan Finley for design and implementation of the program, Jaime Morrison and co-op students, Kelsey Flynn, Rosemary Nguyen, and Dylan Conover for their assistance. We also thank Drs. Espen Rimstad and Øystein Wessel for providing feedback on the manuscript, and the anonymous reviewers for their valuable comments and suggestions to improve the quality of the paper.

# Author Contributions

KMM and ED conceived and designed the study. ED and HWF conducted the histopathology. ED conducted the ISH and immunohistochemistry. KHK, ADS, and SL carried out the molecular analyses. AT provided database and graphical support. OPG provided analytical support associated with the VDD. GM and CS conducted phylogenetic analysis of PRV. ED and KMM co-wrote the manuscript, with input from all authors.

# Competing Interests

No competing interests exist. While samples were provided by the BC aquaculture industry, there was no funding from industry or anti-industry advocates.

# Data Accessibility Statement

PRV genome sequences were deposited to GenBank under accession numbers MH093900 - MH094029

# Tables

**Table 1**. Atlantic salmon collected during an outbreak of HSMI in a longitudinal farm study.

| Group | Fish # | Date | Fork Length (mm) | Total Fish Mass (g) | L-M-D | Epi | End | Heart Score | SkM | Pathologist | PRV (cn/µg) | *Kudoa thyrsites* (cn/µg) | *Desmozoon lepeoptherii* (cn/µg) |
|---|---|---|---|---|---|---|---|---|---|---|---|---|---|
| **1 (CTRL)** | B5586 | 28/08/2013 | 340 | 440 | L | 0 | 0 | 0 | 0 | ED | | | 1.2E+05 |
| | B5589 | 28/08/2013 | 285 | 260 | L | 0 | 0 | 0 | 0 | ED | | 5.9E+07 | 1.2E+06 |
| | B5398 | 28/08/2013 | 350 | 490 | L | 0 | 0 | 0 | 0 | ED | | 4.8E+06 | 6.4E+06 |
| | B5399 | 28/08/2013 | 325 | 365 | L | 0 | 0 | 0 | 0 | ED | | | 1.2E+05 |
| | B5385 | 28/08/2013 | 350 | 465 | L | 0 | 0 | 0 | 0 | ED | | | 1.7E+03 |
| | B5400 | 28/08/2013 | 350 | 550 | L | 0 | 0 | 0 | 0 | ED | | | 4.8E+05 |
| **2 (PRV+/HSMI-)** | B5389 | 28/08/2013 | 350 | 470 | L | 0 | 0 | 0 | 0 | ED | 8.3E+04 | | 3.5E+04 |
| | B5760 | 11/09/2013 | 365 | 590 | L | 0 | 0 | 0 | 0 | ED | 7.7E+05 | | 2.8E+03 |
| | B5903* | 09/10/2013 | 385 | 695 | L | 0 | 0 | 0 | 0 | ED | 5.5E+04 | | 1.1E+04 |
| | B5940 | 09/10/2013 | 385 | 695 | L | 0 | 0 | 0 | 0 | ED | 3.9E+05 | 9.3E+04 | 1.4E+06 |
| | B7274 | 23/10/2013 | 350 | 580 | L | 0 | 0 | 0 | 0 | ED | 2.9E+05 | | 1.6E+06 |
| | B7283 | 23/10/2013 | 300 | 330 | L | 0 | 0 | 0 | 0 | ED | 4.2E+05 | 3.4E+05 | 2.9E+05 |
| **3 (Developing)** | B5392 | 28/08/2013 | 350 | 510 | L | 2 | 1 | 3 | 0 | ED | 1.1E+03 | 2.8E+07 | 1.3E+05 |
| | B5597 | 28/08/2013 | 260 | 230 | D | 1 | 0.5 | 1.5 | 0 | ED | 1.9E+04 | 1.1E+03 | 1.1E+04 |
| | B5910* | 09/10/2013 | 400 | 710 | L | 1 | 0.5 | 1.5 | 0 | ED | 2.3E+05 | | 1.9E+06 |
| | B5933* | 09/10/2013 | 350 | 465 | L | 2 | 1 | 3 | 2 | ED | 1.1E+05 | 5.9E+05 | 4.3E+06 |
| | B5937 | 09/10/2013 | 390 | 675 | L | 2 | 1 | 3 | 1 | ED | 1.8E+05 | 3.3E+05 | 3.2E+07 |
| | B7273 | 23/10/2013 | 360 | 655 | L | 2 | 1 | 3 | 0 | ED | 3.7E+05 | 7.0E+02 | 1.6E+07 |
| **4 (HSMI)** | B5690 | 11/09/2013 | 323 | 350 | D | 3 | 3 | 6 | 3 | ED/HWF | 4.6E+05 | 2.5E+06 | 2.5E+04 |
| | B5688 | 11/09/2013 | 339 | 430 | D | 2 | 2 | 4 | 2 | ED/HWF | 6.3E+04 | | 2.3E+05 |
| | B7293 | 23/10/2013 | 425 | 985 | D | 3 | 3 | 6 | 2 | ED | 5.3E+03 | | 8.0E+04 |
| | B7483 | 05/11/2013 | 450 | 1200 | D | 3 | 3 | 6 | 3 | ED/HWF | 3.1E+04 | | 4.5E+06 |
| | B7474 | 05/11/2013 | 405 | 775 | M | 3 | 3 | 6 | 3 | ED/HWF | 1.4E+04 | | 7.1E+04 |
| | B7473 | 05/11/2013 | 375 | 745 | M | 3 | 3 | 6 | 2 | ED/HWF | 3.7E+04 | 4.3E+05 | 4.7E+05 |

***Note**: Under the "L-M-D" column, "L" was live-, "M" was moribund-, and "D" was dead-sampled. HSMI scoring system includes inflammatory lesions in the epicardium (Epi), endocardium (End) and skeletal muscle (SkM). Also included are the infectious agents detected in the heart of each fish ("cn/µg"="copy number per µg of nucleic acids"). Samples marked with "*" indicated fish tested for infectious agents by pooled tissues (heart, liver, kidney, brain and gills).*

**Table 2**. Distribution of Chinook salmon from the DFO audit program used to study PRV-related jaundice/anemia disease progression.

| | | | | | | | **Viruses (cn/µg)** | | **Bacteria (cn/µg)** | | | | **Parasites (cn/µg)** | | | |
|---|---|---|---|---|---|---|---|---|---|---|---|---|---|---|---|---|
| **Group** | **Fish #** | **Farm code** | **Year** | **Quarter** | **VDD PCA** | **Viral State** | **PRV** | **ENV** | ***P. sal*** | ***R. sal*** | ***A. sal*** | ***V. ang*** | ***L. sal*** | ***D. lep*** | ***P. pse*** | **Diagnosis** |
| **1 (PRV+)** | G572 | P.3-103 | 2013 | 3 | 2.902 | | 7.4E+04 | | | 1.1E+02 | | 3.8E+02 | | 1.1E+02 | 4.4E+04 | nephrocalcinosis |
| | G939 | P.3-166 | 2013 | 4 | 4.820 | | 1.9E+04 | | | | | 1.7E+02 | 1.5E+02 | 5.5E+02 | 3.8E+05 | branchitis |
| | G869 | P.2-99 | 2013 | 3 | 4.263 | | 7.9E+03 | | 6.7E+07 | | | | 3.4E+04 | 9.9E+05 | 1.0E+03 | rickettsiosis |
| **2 (PRV+/VDD+)** | G459 | P.2-3 | 2013 | 1 | -13.308 | VDD | 1.0E+06 | | | | | | | 4.4E+02 | 3.6E+02 | BKD |
| | G363 | P.2-3 | 2013 | 1 | -18.825 | VDD | 5.2E+05 | | | 2.2E+04 | | | 1.0E+06 | 2.2E+02 | | IHN-like |
| | G772 | P.2-99 | 2013 | 3 | -13.475 | VDD | 2.2E+05 | 1.4E+03 | | | | | 6.3E+02 | 4.7E+04 | 5.8E+03 | rickettsiosis |
| | G722 | P.3-8 | 2011 | 4 | -12.240 | VDD | 2.5E+04 | 1.4E+03 | | | | 6.8E+05 | | | 8.9E+05 | vibriosis |
| | G460 | P.2-3 | 2013 | 1 | -10.928 | VDD | 1.1E+06 | | | | | | | 1.6E+06 | | anemia |
| **3 (PRV+/VDD+ jaundice/ anemia)** | G364 | P.2-3 | 2013 | 1 | -11.256 | VDD | 5.1E+05 | | | | | | 2.8E+02 | 3.9E+02 | 5.4E+02 | jaundice/anemia |
| | G579 | P.2-1 | 2011 | 2 | -9.508 | VDD | 1.2E+07 | 3.5E+05 | | | | | 2.9E+06 | | | jaundice |
| | G578 | P.2-1 | 2011 | 2 | -15.870 | VDD | 5.9E+06 | | | | | | | | | jaundice/anemia |
| | G481 | P.2-1 | 2011 | 2 | -13.101 | VDD | 1.9E+06 | | | | | | 5.3E+04 | 2.5E+02 | | jaundice/anemia |
| | G417 | P.3-120 | 2012 | 3 | -15.644 | VDD | 8.5E+05 | 8.0E+04 | | | | | 9.2E+06 | 1.1E+02 | 9.3E+03 | jaundice/anemia |
| | G172 | P.2-3 | 2013 | 1 | -10.282 | VDD | 2.5E+05 | | | | | | 1.9E+03 | 2.0E+03 | 2.6E+02 | jaundice/anemia |
| | G563 | P.2-95 | 2013 | 2 | -17.639 | VDD | 3.5E+05 | 3.9E+04 | | 8.1E+01 | | | | 2.0E+04 | | jaundice |
| | G287 | P.2-95 | 2013 | 2 | -14.137 | VDD | 2.6E+05 | 9.8E+03 | 1.0E+02 | 5.5E+02 | | | | 2.5E+05 | 1.3E+03 | jaundice |
| | G673 | P.2-1 | 2011 | 2 | -10.108 | VDD | 1.6E+05 | | | | | | | | | jaundice |

***Note****: Samples were divided up into three developmental stages, including a pre-disease state whereby fish contained high PRV loads but no evidence of a viral disease state (VDD) or histological lesions consistent with jaundice/anemia, high PRV loads and VDD state, but not diagnosed with jaundice/anemia, and high PRV loads, VDD state, and diagnosed with jaundice/anemia. Shown are the farm regions, year and quarter within which samples were collected and infectious agents detected ("cn/µg" ="copy number per µg of nucleic acids"). The first three digits of the Farm codes indicate regions, with P.2 being farms on the west side of Vancouver Island and P.3 being farms on the east coast of Vancouver Island (Discovery Islands through Broughton (Fig. S1). Acronyms for the infectious agents are: P. sal=Piscirickettsia salmonis, R. sal=Renibacterium salmoninarum, A. sal=Aeromonas salmonicida, V. ang=Vibrio anguillarum, L. sal=Loma salmonae, D. lep= Desmozoon lepeoptherii, P. pse=Parvicapsula pseudobranchicola*

**Table 3**. Histopathology scoring across all tissues of Atlantic salmon over the developmental pathway of HSMI, and Chinook salmon over the developmental pathway of jaundice/anemia.

**Atlantic salmon**

| Group | Fish # | Heart | | Liver | | Spleen | | | Kidney | | | | | Brain | | Gills | | | Muscle | Original Diagnosis | Co-infection agents |
|---|---|---|---|---|---|---|---|---|---|---|---|---|---|---|---|---|---|---|---|---|---|
| | | Epi | End | Nec | Itis | Con | Ell nec | Wpu itis | Itis | Osis | Int pla | Nec | Glo itis | Gli | Con | Itis | Con | Pro | Itis | | |
| 2 (PRV+/HSMI-) | B5389 | 0 | 0 | 0 | 0 | 0 | 0 | 0 | 0 | 0 | 0 | 0 | 0 | 0 | 0 | 0 | 0 | 0 | 0 | | |
| | B5760 | 0 | 0 | 0 | 0 | 0 | 0 | 0 | 0 | 0 | 0 | 0 | 0 | 0 | 0 | 0 | 0 | 0 | 0 | | |
| | B5903 | 0 | 0 | 0 | 0 | 0 | 0 | 0 | 0 | 0 | 0 | 0 | 0 | 0 | 0 | 0 | 0 | 0 | 0 | | |
| | B5940 | 0 | 0 | 0 | 0 | 0 | 0 | 0 | 0 | 0 | 0 | 0 | 0 | 0 | 0 | 0 | 0 | 0 | 0 | | |
| | B7274 | 0 | 0 | 0 | 0 | 0 | 0 | 0 | 0 | 0 | 0 | 0 | 0 | 0 | 0 | 0 | 0 | 0 | 0 | | |
| | B7283 | 0 | 0 | 0 | 0 | 0 | 0 | 0 | 0 | 0 | 0 | 0 | 0 | 0 | 0 | 0 | 0 | 0 | 0 | | |
| 3 (Developing) | B5392 | **2** | **1** | 0 | 0 | 0 | 0 | **1** | 0 | 0 | 0 | 0 | 0 | 0 | 0 | 0 | 0 | 0 | 0 | | |
| | B5597 | **1** | **0.5** | 0 | 0 | 0 | 0 | 0 | 0 | 0 | 0 | 0 | 0 | 0 | 0 | 0 | 0 | 0 | 0 | | |
| | B5910 | **1** | **0.5** | 0 | 0 | 0 | 0 | 0 | 0 | 0 | 0 | 0 | 0 | 0 | 0 | 0 | 0 | 0 | 0 | | |
| | B5933 | **2** | **1** | 0 | 0 | 0 | 0 | **1** | 0 | 0 | **1** | 0 | 0 | 0 | 0 | **1** | 0 | 0 | **2** | | |
| | B5937 | **2** | **1** | 0 | 0 | 0 | 0 | **1** | 0 | 0 | **1** | 0 | 0 | 0 | 0 | 0 | 0 | 0 | **1** | | |
| | B7273 | **2** | **1** | 0 | 0 | 0 | 0 | 0 | 0 | 0 | 0 | 0 | 0 | 0 | 0 | 0 | 0 | 0 | 0 | | |
| 4 (HSMI) | B5690 | **3** | **3** | **2** | 0 | 0 | **1** | 0 | 0 | 0 | 0 | 0 | 0 | 0 | 0 | **2** | 0 | 0 | **3** | HSMI | |
| | B5688 | **2** | **2** | **2** | 0 | **1** | **2** | **1** | 0 | 0 | **1** | 0 | 0 | 0 | **1** | 0 | 0 | 0 | **2** | HSMI | |
| | B7293 | **3** | **3** | **1** | 0 | 0 | **1** | **1** | 0 | 0 | **1** | 0 | 0 | 0 | 0 | 0 | 0 | 0 | **2** | HSMI | |
| | B7483 | **2** | **3** | **2** | 0 | **2** | **2** | **1** | 0 | 0 | **1** | 0 | 0 | 0 | 0 | 0 | 0 | 0 | **3** | HSMI | |
| | B7474 | **2** | **2** | **2** | 0 | **1** | **1** | 0 | 0 | 0 | 0 | 0 | 0 | 0 | 0 | 0 | 0 | 0 | **3** | HSMI | |
| | B7473 | **1** | **3** | **2** | 0 | **1** | **2** | 0 | 0 | 0 | 0 | 0 | 0 | 0 | 0 | 0 | 0 | 0 | **3** | HSMI | |

**Chinook salmon**

| Group | Fish # | Epi | End | Nec | Itis | Con | Ell nec | Wpu itis | Itis | Osis | Int pla | Nec | Glo itis | Gli | Con | Itis | Con | Pro | Itis | Original Diagnosis | Co-infection agents |
|---|---|---|---|---|---|---|---|---|---|---|---|---|---|---|---|---|---|---|---|---|---|
| 1 (PRV+) | G572 | 0 | 0 | 0 | 0 | 0 | 0 | 0 | 0 | 0 | **1** | 0 | 0 | 0 | 0 | 0 | 0 | 0 | 0 | nephrocalcinosis | |
| | G939 | 0 | 0 | 0 | 0 | 0 | 0 | 0 | 0 | 0 | **2** | 0 | 0 | 0 | 0 | **1** | 0 | **1** | 0 | branchitis | *P. pseudobranchicola* |
| | G869 | 0 | 0 | 0 | **3** | 0 | 0 | 0 | 0 | **2** | **2** | 0 | 0 | 0 | 0 | 0 | 0 | 0 | 0 | rickettsiosis | *P. salmonis* |
| 2 (PRV+/VDD+) | G459 | 0 | 0 | **2** | 0 | **2** | **2** | 0 | 0 | **2** | **2** | 0 | **2** | 0 | 0 | **1** | 0 | 0 | 0 | BKD | *R. salmoninarum* |
| | G363 | 0 | 0 | **3** | 0 | 0 | **3** | 0 | 0 | 0 | **2** | **2** | 0 | 0 | 0 | 0 | 0 | 0 | 0 | IHN-like? | *L. salmonae* |
| | G772 | 0 | 0 | **2** | 0 | **1** | **1** | **1** | 0 | **1** | **3** | 0 | 0 | **1** | 0 | 0 | 0 | 0 | 0 | rickettsiosis | NO *P. salmonis* |
| | G722 | 0 | 0 | 0 | **2** | 0 | 0 | **2** | 0 | 0 | 0 | 0 | 0 | 0 | 0 | 0 | 0 | 0 | 0 | vibriosis | *V. anguillarum* |
| | G460 | 0 | **1** | 0 | 0 | **2** | 0 | 0 | 0 | **2** | **1** | 0 | 0 | 0 | 0 | 0 | 0 | **1** | 0 | anemia | |
| 3 (PRV+/VDD+ jaundice/ anemia) | G364 | 0 | **3** | **2** | 0 | **1** | 0 | 0 | 0 | 0 | **2** | 0 | 0 | 0 | 0 | 0 | 0 | 0 | 0 | jaundice/anemia | |
| | G579 | 0 | 0 | **2** | 0 | 0 | **3** | **3** | **2** | 0 | **3** | 0 | 0 | 0 | 0 | 0 | 0 | 0 | 0 | jaundice | ENV, *L. salmonae* |
| | G578 | 0 | **3** | **2** | **3** | **1** | **1** | **3** | **3** | 0 | **2** | 0 | 0 | 0 | 0 | 0 | 0 | 0 | 0 | jaundice/anemia | |
| | G481 | 0 | 0 | **0*** | 0 | **2** | 0 | 0 | 0 | **2** | **2** | 0 | 0 | 0 | 0 | 0 | 0 | 0 | 0 | jaundice/anemia | |
| | G417 | 0 | 0 | 0 | 0 | 0 | 0 | **2** | **2** | 0 | 0 | 0 | 0 | 0 | 0 | **2** | 0 | 0 | 0 | jaundice/anemia | *L. salmonae* |
| | G172 | 0 | 0 | 0 | 0 | 0 | 0 | 0 | 0 | **2** | **3** | 0 | 0 | 0 | 0 | 0 | 0 | 0 | 0 | jaundice/anemia | |
| | G563 | 0 | 0 | **1** | 0 | **1** | **2** | 0 | **1** | 0 | **2** | 0 | 0 | 0 | 0 | 0 | 0 | 0 | 0 | jaundice | |
| | G287 | 0 | 0 | **0*** | 0 | 0 | 0 | **2** | 0 | 0 | **2** | 0 | 0 | **1** | 0 | 0 | **3** | 0 | 0 | jaundice | |
| | G673 | 0 | **2** | **1** | 0 | 0 | **1** | 0 | 0 | **3** | **2** | 0 | 0 | 0 | 0 | 0 | 0 | 0 | 0 | jaundice | |

***Note:*** *Lesion scores different than zero are highlighted in bold. Abbreviations for lesions are as follows: Epi=epicarditis, End=endocarditis, Con=congestion, Nec=necrosis, Itis=inflammation, Ell nec=ellipsoid necrosis, Wpu Itis=white pulp inflammation, Osis=tubule necrosis, Int pla=interstitial tissue hyperplasia, Glo Itis=glomeruli inflammation, Gli=gliosis, Pro=proliferation. Other lesions (i.e. liver congestion, spleen capsule proliferation, pancreatitis, enteritis, encephalitis, cerebral malacia and cerebral microsporidia) have been assessed, but did not occur in any fish, therefore have not been included in the table. Two fish (marked with*) had no necrotic areas in the liver, but virtually all hepatocytes had varying degrees of vacuolar degeneration, characterized by a foamy cytoplasm.*